\documentclass[floats,floatfix,showpacs,preprintnumbers,amssymb,prd,twocolumn,superscriptaddress,nofootinbib,nolongbibliography,reprint]{revtex4-1}

\usepackage{amssymb,amsmath,verbatim,mathtools,needspace,enumitem,etoolbox,graphicx,physics,microtype,afterpage,xspace,tabularx,lmodern,multirow}
\usepackage{gensymb}
\usepackage{appendix}
\usepackage{tensor}
\usepackage[normalem]{ulem}
\usepackage[dvipsnames, usenames]{xcolor}
\usepackage{xr-hyper}
\definecolor{linkcolor}{rgb}{0.0,0.3,0.5}
\usepackage[unicode, colorlinks=true, linkcolor=linkcolor, citecolor=linkcolor, filecolor=linkcolor, urlcolor=linkcolor, linktocpage, breaklinks]{hyperref}
\usepackage[all]{hypcap}
\usepackage[T1]{fontenc}
\usepackage[utf8]{inputenc}
\usepackage[usenames,dvipsnames]{xcolor}
\hypersetup{colorlinks=true,citecolor=magenta,linkcolor=violet,urlcolor=magenta}

\definecolor{romared}{RGB}{142,0,28}

\newcommand{\be}{\begin{equation}}
\newcommand{\ee}{\end{equation}}

\def\be{\begin{equation}}
\def\ee{\end{equation}}
\newcommand{\beq}{\begin{eqnarray}}
\newcommand{\eeq}{\end{eqnarray}}

\usepackage{makecell}
\usepackage{soul}

\newcolumntype{Y}{>{\centering\arraybackslash}X}

\makeatletter
\newcommand*{\addFileDependency}[1]{
  \typeout{(#1)}
  \@addtofilelist{#1}
  \IfFileExists{#1}{}{\typeout{No file #1.}}
}
\makeatother

\vspace{1cm}

\begin{document}

\title{Black Hole Perturbations and Electric-Magnetic Duality}
\author{David Pere\~niguez}
\affiliation{Niels Bohr International Academy, Niels Bohr Institute, Blegdamsvej 17, 2100 Copenhagen, Denmark}
%
%
%
%
\begin{abstract} 
Black holes can be electromagnetically charged, or carry vector charge from new fundamental fields. Their response to small fluctuations is of paramount importance to study gravitational wave generation. However, the usual even and odd sectors of gravitoelectromagnetic waves couple if the black hole is magnetically charged, a fact that complicates significantly the perturbative approach. In this paper, perturbation theory based on harmonic expansion is extended to have manifest invariance under electric-magnetic duality. As a result, the equations decouple into two generalised even and odd sectors, each governed by master wave equations that include the most general coupling to a dyonic source. These can be used to compute, in a simple manner, the gravitational and electromagnetic radiation emitted in the interaction of the most general spherically symmetric black holes of the Einstein--Maxwell theory with electromagnetically charged matter.  
\end{abstract}

\maketitle 


\section{Introduction}

According to General Relativity (GR), all quiescent black holes (BHs) in our universe are uniquely described by their mass, angular momentum, electric and magnetic charge~\cite{Carter:1971zc,PhysRevLett.34.905,Ruffini:1971bza,Hawking:1973uf}. The deep implications of this result makes charged BHs an appealing class of compact objects that, as a matter of fact, have captured the interest of theoretical astrophysicists for decades~\cite{Zerilli:1974ai,Johnston:1974vf,PhysRevD.10.1057,Gerlach:1979rw,Gerlach:1980tx,Chandrasekhar1979OnTM}. In particular, they are a well-defined extension of the vacuum Kerr BH within a realistic theory, and thus constitute an ideal paradigm for multi-messenger astronomy. While assuming that BHs are neutral is a reasonable and well-motivated simplification (because of friction with interstellar medium and Schwinger pair-creation of electric charges \cite{Gibbons:1975kk}, and from the lack of evidence of magnetic charges in laboratory experiments and cosmic ray observations), the desirable scientific programme is to perform an analysis of the most general BHs allowed in a given theory first, and then constrain their parameters by contrasting with observations (in fact, there are well known astrophysical mechanisms through which BHs acquire charge, even though small \cite{Wald:1974np,Beskin:2000qe,Palenzuela:2011es}). This is even more important nowadays given the current stage of Gravitational Wave (GW) astronomy, which allows unprecedented tests of the strong field regime of gravitation.

Charged BHs do also provide a unique opportunity in searches of new physics from GWs. First conjectured by Dirac \cite{Dirac:1931kp}, magnetic monopoles could have been produced in the early Universe, as robustly predicted by GUTs (the so-called primordial monopoles)~\cite{Preskill:1984gd}, and it is not unreasonable that BHs formed at that time could have accreted some net amount of magnetic charge, or that were formed directly from the collapse of the latter \cite{Gibbons:1990um}. Since magnetic monopoles are more stable to Schwinger pair decay, these magnetic BHs would have Hawking-evaporated until reaching extremality and could correspond to a fraction of the dark matter content in the Universe, and are also an interesting alternative solution to the monopole problem in cosmology \cite{Bai:2019zcd,Stojkovic:2004hz,PhysRevD.26.1296,Maldacena:2020skw,Kritos:2021nsf}. In addition, strong magnetic fields such as those in the vicinity of an extremal magnetic BH would have remarkable consequences on Standard Model fields \cite{Maldacena:2020skw}. Finally, BH charge could also be due to milicharged dark matter and hidden vector fields, as invoked by beyond-the-Standard-Model physics (including some dark matter models), which can easily circumvent standard discharge mechanisms~\cite{DeRujula:1989fe,Perl:1997nd,Holdom:1985ag,Sigurdson:2004zp,Davidson:2000hf,McDermott:2010pa,Cardoso:2016olt,Khalil:2018aaj,Bai:2019zcd,Gupta:2021rod,Kritos:2021nsf}. 

Deriving GW bounds for charged BHs is thus an interesting problem that should complement current constraints from other perspectives \cite{Zhang:2023zmb,Kobayashi:2023ryr}. In isolation, it is possible to constrain the ``total charge'' of a BH given by the duality-invariant quantity $\sqrt{Q^{2}+P^{2}}$, where $Q$ and $P$ are respectively the electric and magnetic charge. Thus, by electric-magnetic duality it suffices to restrict to the purely electric case in e.g.~ringdown and stability analysis \cite{Zerilli:1974ai,Johnston:1974vf,PhysRevD.10.1057,Gerlach:1979rw,Gerlach:1980tx,Chandrasekhar1979OnTM,Carullo:2021oxn,Dias:2021yju}. However, in interaction with other charges (e.g.~during accretion of matter or in the inspiral phase of a merger) there are effects via which $P$ and $Q$ can be constrained separately \cite{HGed,Garfinkle:1990zx,Dyson:2023ujk}. While there is a large body of work about electric BHs accreting electrically charged matter, from extreme mass-ratio mergers to comparable-mass BH coalesces \cite{Gupta:2021rod,Zerilli:1974ai,Johnston:1974vf,Bozzola:2020mjx,Zilhao:2012gp,Zilhao:2013nda,Liebling:2016orx}, much less is known about more general scenarios in which a dyonic BH (a BH with both electric and magnetic charge) interacts with charged matter. These events are not related via electric-magnetic duality to purely electric ones, and could lead to interesting novel constraints on the BH and matter parameters, which should be compatible with those obtained from tests in isolation. 

One natural approach to study these systems is to take the Newtonian limit, where motion is non-relativistic and matter is modelled as dyonic point charges \cite{Liu:2020cds,Liu:2020vsy,Liu:2020bag,Liu:2022wtq,Chen:2022qvg}. However, from the perspective of GW astronomy it is imperative to derive theoretical predictions that include both strong field and relativistic effects. In mergers with extreme mass ratios, which are of much relevance for low frequency GW detectors such as LISA, perturbation theory provides very accurate results and its input is crucial for the construction of waveform templates. However, a perturbative treatment gets complicated by the well-known fact that the usual even and odd sectors of gravitoelectromagnetic waves couple if a BH is magnetically charged \cite{Gerlach:1979rw,Gerlach:1980tx,Kodama:2003kk}.

Here, this problem is fixed by devising an harmonic approach to perturbation theory that is manifestly invariant under the electric-magnetic duality transformations and the gauge symmetry of the linear theory. As a result, the linearised Einstein--Maxwell equations decouple into two generalised even and odd sectors, and are governed by master wave equations that include the coupling to the most general dyonic matter sources.

The paper is organised as follows. In Section \ref{EMduality} we briefly review electric-magnetic duality and introduce the most general spherically symmetric BHs of the Einstein--Maxwell theory. Next, in Section \ref{LinFluc} we first introduce a covariant and gauge-invariant formalism to describe fluctuations of spherically symmetric spacetimes where the energy-momentum tensor (both the background and the fluctuations) is completely general (Section \ref{S1}). Then, we specialise the equations to the Einstein--Maxwell theory in a way that electric-magnetic self-duality is manifest, and show that the linearised equations decouple into two generalised even and odd sectors (Section \ref{S2}). Finally, we derive decoupled master wave equations governing the dynamics of each sector (Section \ref{S3}). We conclude in Section \ref{Disc} discussing our results and future research directions. 

\section{Dyonic Black Holes and Electric-Magnetic Duality}\label{EMduality}
The Einstein--Maxwell theory coupled to additional matter is governed by the equations
\begin{align}\label{E1}
G_{\mu\nu}+\Lambda g_{\mu\nu}&=2\mathcal{F}_{\mu\alpha}\mathcal{F}_{\nu}^{\ \alpha}-\frac{1}{2}g_{\mu\nu}\mathcal{F}^{2}+S_{\mu\nu}\, ,\\ \label{Mm1} 
d \mathcal{F}&=-4\pi \star J_{(m)}\, ,\\ \label{Mm2}
 d\star \mathcal{F}&=-4\pi \star J_{(e)} \,,
 \end{align}
where a cosmological constant $\Lambda$ is included for completeness, and throughout this work we use geometric units $G=c=1$. Here, $\mathcal{F}_{\mu\nu}$ is the Maxwell field strength and $J_{(e)}^{\mu}, J_{(m)}^{\mu}$ and ${S}_{\mu\nu}$ are, respectively, the electric and magnetic currents and the energy-momentum tensor associated to the additional matter. Consistency requires on-shell conservation of the currents and the total energy momentum tensor (the right hand side of \eqref{E1}), that is,\footnote{Although not completely immediate, it is a standard exercise to write $\nabla^{\mu}(S_{\mu\nu}+2\mathcal{F}_{\mu\alpha}\mathcal{F}_{\nu}^{\ \alpha}-\frac{1}{2}g_{\mu\nu}\mathcal{F}^{2})=0$ in the form \eqref{cons3}, assuming the Maxwell equations \eqref{Mm1}-\eqref{Mm2}.}
\begin{align}\label{cons1}
d\star J_{(e)}&=0\, ,\\ \label{cons2}
d\star J_{(m)}&=0\, , \\  \label{cons3}
\nabla^{\mu}S_{\mu\nu}&=-8\pi\left(J^{\alpha}_{(m)}\left(\star \mathcal{F}_{\nu\alpha}\right)-J^{\alpha}_{(e)} \mathcal{F}_{\nu\alpha}\right)\, .
\end{align}
These equations must hold regardless of the kind of matter considered. The idea now is to cast equations \eqref{E1}-\eqref{cons3} in a form that electric-magnetic self-duality is manifest. To that end, we introduce the complex field strength and current
\begin{equation}
\begin{aligned} \label{complexF}
\bold{F}\equiv \mathcal{F}-i\star\mathcal{F}\,, \ \ \ \ \ \ \bold{J}\equiv J_{(m)}-i J_{(e)}\,,
\end{aligned}
\end{equation}
in terms of which the action of an electric-mangetic duality transformation is
\begin{align}\label{phasetrafo}
\bold{F}&\mapsto e^{i\alpha}\bold{F}\,, \ \ \ \bold{J}\mapsto e^{i\alpha}\bold{J}\,, \ \ \ \alpha\in\mathbb{R}\, .
\end{align}
This is nothing but an $SO(2)$ transformation of the field strength and the current, while the spacetime metric $g_{\mu\nu}$ and the matter energy-momentum tensor $S_{\mu\nu}$ are left invariant (a paradigmatic example where this symmetry is realised is the Einstein--Maxwell theory coupled to a dyonic point particle \cite{Carter:1973rla}). Now, equations \eqref{E1}-\eqref{Mm2} and the conservation laws  \eqref{cons1}-\eqref{cons3} take the form
\begin{align}\label{einstein}
G_{\mu\nu}+\Lambda g_{\mu\nu}&=\bold{F}_{\mu\alpha}\bar{\bold{F}}_{\nu}^{\ \alpha}+{S}_{\mu\nu}\,, \\  \label{max1}
d \bold{F}&=\star \bold{J} \, ,\\  \label{max2}
\bold{F}+i\star \bold{F}&=0\,, 
\end{align}
and 
\begin{align}\label{consC1}
 d\star \bold{J}&=0\,, \\ \label{consC2}
 \nabla^{\mu}{S}_{\mu\nu}&=-4\pi i\left(\bar{\bold{J}}^{\alpha}\bold{F}_{\nu\alpha}-\bold{J}^{\alpha}\bar{\bold{F}}_{\nu\alpha}\right)\, ,
\end{align}
where the bar denotes complex conjugation, so they are manifestly invariant under the transformation \eqref{phasetrafo}. It is worth noticing that, although electric-magnetic duality transformations may not be defined for certain classes of additional matter fields (e.g.~an abelian-Higgs model), this does not obstruct by any means the possibility of working in terms of the variables \eqref{complexF}, which in any case must be subject to equations \eqref{einstein}-\eqref{max2}, and the conservation laws \eqref{consC1} and \eqref{consC2}.

We will focus on the most general electrovacuum ($S_{\mu\nu}=0$ and  $\bold{J}=0$) spherically symmetric BH solution of equations \eqref{einstein}-\eqref{max2}. This is the dyonic Reissner--Nordström--(Anti)de Sitter (RN(A)dS) BH \cite{Reissner,Nordstrom}, which in Schwarzschild coordinates reads
\begin{equation}
\begin{aligned}\label{RN}
ds^{2}&=-f(r)dt^{2}+\frac{dr^{2}}{f(r)}+r^{2}\left(d\theta^{2}+\sin^{2}\theta d\phi^{2}\right)\, ,\\
\bold{F}&=-i\frac{\bold{C}}{r^{2}}dt\wedge dr+\bold{C}\sin\theta d\theta\wedge d\phi\, ,
\end{aligned}
\end{equation}
where 
\begin{align}
f(r)=1-\frac{2M}{r}-\frac{\Lambda}{3}r^{2}+\frac{\bar{\bold{C}}\bold{C}}{r^{2}}  \,.
\end{align}
$M$ is the BH mass and $\bold{C}$ is a combination of its electric and magnetic charges, $Q$ and $P$ respectively, defined as
\begin{equation}
\bold{C}\equiv\frac{1}{4\pi}\int_{S^{2}} \bold{F}=P-iQ\, , \\
\end{equation}
where $S^{2}$ is any sphere that encloses the BH. Finally, we notice that the action of a duality rotation on the solution \eqref{RN} is simply $\bold{C}\mapsto e^{i\alpha}\bold{C}$.

\section{Linear Fluctuations}\label{LinFluc}

In this section we consider linear fluctuations of the spacetime \eqref{RN}. First, in Section~\ref{S1} we establish a covariant and gauge-invariant formalism to treat perturbations on background spacetimes that are spherically symmetric, allowing for the most general energy-momentum tensor both at the background and linear levels (the formalism is an extension of \cite{Gerlach:1979rw,Gerlach:1980tx,Martel:2005ir} to include an arbitrary energy-momentum tensor, and is inspired in part by the approach of \cite{Ishibashi:2011ws}). In Section~\ref{S2} we specialise the previous general equations to the Einstein--Maxwell case, in such a way that duality-invariance is manifest and the equations decouple into two generalised even and odd sectors, which are defined irrespective of the charge configurations of the background and the sources. Finally, in Section~\ref{S3} we derive decoupled master wave equations that govern each sector and include the most general dyonic source terms.

\subsection{Einstein Equations with General Matter}\label{S1}

For the sake of generality, here we consider spacetimes of dimension $N+2$, whose background metric and energy-momentum tensor have the form
\begin{align}\label{linel}
ds^{2}&=g_{ab}(y)dy^{a}dy^{b}+r^{2}(y)\Omega_{AB}(z)dz^{A}dz^{B}\,, \\ \notag \\ \label{T}
T&=T_{ab}(y)dy^{a}dy^{b}+r^{2}(y)\mathcal{T}(y)\Omega_{AB}dz^{A}dz^{B}\, ,
\end{align}
which are defined on a manifold with structure $M=\mathcal{N}^{N}\times \mathbb{S}^{2}$. $g_{ab}(y)$ and $r^{2}(y)$ are a Lorentizan metric and a positive function in the manifold $\mathcal{N}^{N}$, which is parametrised by the coordinates $y^{a}$ (with $a=1,..., N$). Similarly, $T_{ab}(y)$ and $\mathcal{T}(y)$ are a symmetric tensor and a function in $\mathcal{N}^{N}$. The coordinates $z^{A}$ (with $A=1,2$) parametrise the unit round 2-sphere $\mathbb{S}^{2}$ with metric $\Omega_{AB}(z)$ (note that no assumption is made about the choice of neither $y^{a}$ nor $z^{A}$). There is a large list of important spacetimes that belong to the class \eqref{linel}, from rotating BHs in higher dimensions to four-dimensional spacetimes of the form \eqref{RN}, which are the relevant ones in this work. We shall fix some conventions at this point. Greek characters are reserved for spacetime indices of the total manifold $M$ (e.g.~$\mu=1,...,N+2$), and the total spacetime metric is denoted by a hat $\hat{g}_{\mu\nu}$, as well as its associated covariant derivative and curvature tensor $\hat{\nabla}$ and $\hat{R}^{\mu}_{\ \nu\rho\sigma}$. Indices are raised and lowered with the metrics $g_{ab}$ and $\Omega_{AB}$, whose covariant derivatives  and curvature tensors are denoted $\nabla, R^{a}_{\ bcd}$ and $D,\mathcal{R}^{A}_{\ BCD}$, respectively. Finally, it will be useful to introduce the 1-form $r_{a}\equiv \nabla_{a}r$, the function $H\equiv r^{a}r_{a}-1$, and the metric volume forms $\varepsilon_{ab}$ and $\epsilon_{AB}$ associated to $g_{ab}$ and $\Omega_{AB}$, respectively.

In this terminology, Einstein's equations and the conservation law satisfied by the background \eqref{linel}-\eqref{T} read
\begin{align}
G_{ab}+\left(\frac{H}{r^{2}}+\Lambda\right)g_{ab}-\frac{2}{r}\left(\nabla_{a}r_{b}-\nabla_{c}r^{c} g_{ab}\right)&=T_{ab}\, , \\
\frac{\nabla_{a}r^{a}}{r}-\frac{R}{2}+\Lambda&=\mathcal{T} \,, \\
\nabla^{b}\left(r^{2}T_{ba}\right)-2r \mathcal{T} r_{a}&=0\, .
\end{align}

First order dynamical deviations from the background \eqref{linel}-\eqref{T} are described by the metric and energy-momentum fluctuations $h_{\mu\nu}$ and $\delta T_{\mu\nu}$. These are subject to the linearised Einstein equations and to the conservation of energy and momentum,
\begin{align}\label{linEinst}
&\delta \hat{G}_{\mu\nu}+\Lambda h_{\mu\nu}=\delta T_{\mu\nu}+S_{\mu\nu}\, , \\ \notag \\ \label{emCons}
&\delta\left(\hat{\nabla}^{\mu}T_{\mu\nu}\right)+\hat{\nabla}^{\mu}S_{\mu\nu}=0\, ,
\end{align}
where we are writing the total first order energy-momentum tensor as $\delta T_{\mu\nu}+S_{\mu\nu}$, with $\delta T_{\mu\nu}$ denoting the contribution associated to the fields whose background value is non-zero, while $S_{\mu\nu}$ is the contribution from additional matter fields that vanish on the background, and so correspond to a first order source. Next, we expand the fluctuations in spherical tensor harmonics as
\begin{align}\label{hdec}\notag
h=&\ h^{\ell}_{ab}(y)Y^{\ell}dy^{a}dy^{b}+2\left[h^{\ell}_{a}(y)Z^{\ell}_{A}+j^{\ell}_{a}(y)X^{\ell}_{A}\right]dy^{a}dz^{A} \\
&+\left[j^{\ell} (y)W^{\ell}_{AB}+k^{\ell} (y)U^{\ell}_{AB}+m^{\ell}(y)V^{\ell}_{AB}\right]dz^{A}dz^{B}\, , \\ \notag \\ \notag
\delta T=&\ \theta^{\ell}_{ab}(y)Y^{\ell}dy^{a}dy^{b}+2\left[\theta^{\ell}_{a}(y)Z^{\ell}_{A}+\rho^{\ell}_{a}(y)X^{\ell}_{A}\right]dy^{a}dz^{A} \\ \label{Tdec}
&+\left[\rho^{\ell}(y) W^{\ell}_{AB}+\theta^{\ell}(y) U^{\ell}_{AB}+\sigma^{\ell}(y)V^{\ell}_{AB}\right]dz^{A}dz^{B}\, ,
\end{align}
where $Y^{\ell},Z^{\ell}_{A},U_{AB}^{\ell},V_{AB}^{\ell}$ and $X^{\ell}_{A},W^{\ell}_{AB}$ are the even and odd spherical tensor harmonics, respectively, which are labelled by the usual quantum numbers $\ell=(l,m)$ and summation over repeated $\ell$'s is assumed (although we shall omit writing this index to alleviate the notation). Our conventions in defining the spherical harmonics are given in Appendix \ref{SHandEE}. The even and odd sectors of the fluctuations $h_{\mu\nu}$ and $\delta T_{\mu\nu}$ consist of their components relative to the even and odd spherical harmonics, respectively (e.g.~$h^{\ell}_{ab}(y),h^{\ell}_{a}(y),k^{\ell} (y),m^{\ell}(y)$ form the even sector of $h_{\mu\nu}$ while $j^{\ell}_{a}(y),j^{\ell}(y)$ form the odd one). These components are well defined tensors on $\mathcal{N}^{N}$, and the equations of motion \eqref{linEinst}-\eqref{emCons} reduce to a set of linear PDE's for them. However, the gauge symmetry
\begin{align}
h_{\mu\nu}&\mapsto h_{\mu\nu}-2\hat{\nabla}_{(\mu}\xi_{\nu)}\, , \\ \notag \\
\delta T_{\mu\nu}&\mapsto \delta T_{\mu\nu}-\pounds_{\xi}T_{\mu\nu}\, ,
\end{align}
implies that some of the degrees of freedom are unphysical. One customary approach is to chose a suitable gauge, but here we shall work with gauge-invariant variables that can be constructed systematically as follows. Expanding the gauge parameter $\xi_{\mu}$ in harmonics,
\begin{equation}
\xi=\xi_{a}Ydy^{a}+\left[\xi Z_{A}+\chi  X_{A}\right]dz^{A}\, ,
\end{equation}
it is easy to check that the fluctuation-dependent vector field $\eta[h]=\eta_{a}[h] dy^{a}+\left(\eta[h]Z_{A}+\upsilon[h]X_{A}\right)dz^{A}$, with 
\begin{align}\label{vec1}
\eta_{a}[h]&\equiv -h_{a}+\frac{r^{2}}{2}\nabla_{a}\left(\frac{m}{r^{2}}\right)\, ,\\ \label{vec2}
\eta[h]&\equiv -\frac{m}{2}\, ,\\\label{vec3}
\upsilon[h]&\equiv -\frac{j}{2}\, ,
\end{align}
transforms as
\begin{equation}
\eta_{\mu}[h] \mapsto \eta_{\mu}[h]+\xi_{\mu}\, .
\end{equation}
Then the variables
\begin{align}\label{htilde}
\tilde{h}&\equiv\left(h_{\mu\nu}+2\hat{\nabla}_{(\mu}\eta_{\nu)}\right)dx^{\mu}dx^{\nu}\, , \\ \notag \\
\tilde{\theta}&\equiv\left(\delta T_{\mu\nu}+\pounds_{\eta}T_{\mu\nu}\right)dx^{\mu}dx^{\nu}\, ,
\end{align}
are manifestly gauge-invariant, and we shall work in terms of their harmonic components, denoted
\begin{align}\label{htildeCOMPS}
\tilde{h}=&\tilde{h}_{ab}Ydy^{a}dy^{b}+2\tilde{j}_{a}X_{A}dy^{a}dz^{A}+\tilde{k}U_{AB}dz^{A}dz^{B}\, , \\ \notag \\ \notag
\tilde{\theta}=&\tilde{\theta}_{ab}Ydy^{a}dy^{b}+2\left[\tilde{\theta}_{a}Z_{A}+\tilde{\rho}_{a}X_{A}\right]dy^{a}dz^{A} \\
&+\left[\tilde{\rho}W_{AB}+\tilde{\theta}U_{AB}+ \tilde{\sigma}V_{AB}\right]dz^{A}dz^{B}\, ,  \label{TtildeCOMPS}
\end{align}
whose explicit expression in terms of the original ones \eqref{hdec} and \eqref{Tdec} is given in Appendix \ref{SHandEE}. 

Inserting this expansion into the linearised Einstein's equations \eqref{linEinst} one finds that they decouple into two sets. The first set contains only the even sector of \eqref{htildeCOMPS} and \eqref{TtildeCOMPS}, and reads
\begin{equation}\label{even}
E_{ab}=\tilde{\theta}_{ab}+\Sigma_{ab}\, , \ \ \ E_{a}=\tilde{\theta}_{a}+ \Sigma_{a}\, , \ \ \ E=\tilde{\theta}+\Sigma \, , \ \ \ \mathcal{E}=\sigma+\mathcal{S}\, ,
\end{equation}
where $E_{ab},E_{a},E,\mathcal{E}$ are given in Appendix \ref{SHandEE}, and the source terms are 
\begin{align}
\Sigma_{ab}&\equiv \int d\Omega \bar{Y}^{L}S_{ab}\, ,\\
\Sigma_{a}&\equiv \frac{1}{l(l+1)}\int  d\Omega \bar{Z}^{LA}S_{aA} \, ,\\
\Sigma&\equiv\frac{1}{2}\int d\Omega \bar{U}^{LAB}S_{AB} \, ,\\
\mathcal{S}&\equiv2\frac{(l-2)!}{(l+2)!}\int d\Omega \bar{V}^{LAB}S_{AB} \, .
\end{align}
The other set of equations contains only the odd sector of \eqref{htildeCOMPS} and \eqref{TtildeCOMPS}, it reads
\begin{equation}\label{odd}
O_{a}=\tilde{\rho}_{a}+\Upsilon_{a}\, , \ \ \ O=\tilde{\rho}+\Upsilon\, ,
\end{equation}
where $O_{a},O$ are given in Appendix \ref{SHandEE}, and the source terms are 
\begin{align}
\Upsilon_{a}&\equiv \frac{1}{\lambda^{2}}\int  d\Omega \bar{X}^{LA}S_{aA}\, , \\
\Upsilon&\equiv 2\frac{(l-2)!}{(l+2)!}\int d\Omega \bar{W}^{LAB}S_{AB}\, .
\end{align}
Finally, the total energy-momentum tensor is necessarily conserved on-shell. Thus, Einstein's equations \eqref{even} and \eqref{odd} have to be supplemented with the conservation laws that result from plugging the expansions \eqref{htildeCOMPS} and \eqref{TtildeCOMPS} into \eqref{emCons}. Again, even and odd sectors decouple and the explicit form of the conservation equations is reported in Appendix \ref{SHandEE}. 
\subsection{The Einstein--Maxwell Case}\label{S2}
In this section we consider fluctuations of the dyonic RN(A)dS BH \eqref{RN} in the Einstein--Maxwell theory. These are governed by equations \eqref{einstein}-\eqref{max2} linearised on the background \eqref{RN}, and we shall also include a general dyonic source, with current $\bold{J}$ and energy--momentum tensor $S_{\mu\nu}$ (examples of such sources are a dyonic point charge \cite{Carter:1973rla}, a complex, charged scalar wave \cite{Natario:2016bay}, etc), which should satisfy equations \eqref{consC1}-\eqref{consC2} for consistency. Most of the work concerning the linearisation of Einstein's equations was done in the previous section for general matter fields and sources. Specialising those equations to the matter content of Maxwell's theory requires, first, defining an harmonic expansion for the fluctuation of Maxwell's field, then linearise Maxwell's equations \eqref{max1}-\eqref{max2} and, finally, compute the energy-momentum tensor fluctuations \eqref{TtildeCOMPS} in terms of Maxwell's field. 

Consider first a BH background that is only electrically charged and sources that are purely electric. Then it is enough to expand the perturbed Maxwell vector potential $\delta \mathcal{A}_{\mu}$ in even and odd harmonics, since in that case the even components of the electromagnetic field couple only to the even components of the gravitational one, and similarly for the odd components \cite{Zerilli:1974ai,PhysRevD.10.1057,Chandrasekhar:1985kt}, so the even and odd sectors of the gravitoelectromagnetic fluctuation decouple. If the background BH carries magnetic charge, though, such an approach does not lead to decoupled equations.\footnote{This was noticed in earlier works such as \cite{Gerlach:1979rw,Gerlach:1980tx,Kodama:2003kk}.} To see this, consider electrovacuum fluctuations so all sources are set to zero. Then, proceeding as above and expanding $\delta \mathcal{A}_{\mu}$ in harmonics one finds that it is a mixed combination of even and odd components of the Maxwell field that sources each gravitational sector, thus spoiling the decoupling of the Einstein--Maxwell equations.\footnote{The reason why this happens is that now the background Maxwell field strength contains a term $\sim P \epsilon_{AB} $, where $P$ is the BH magnetic charge and $\epsilon_{AB}$ the volume form in the 2-sphere. Every time this background piece is contracted with an even (odd) vector harmonic $Z_{A}$ ($X_{A}$) one gets back an odd (even) one, since $X_{A}=\epsilon_{AB}Z^{B}$, causing the above mentioned mixing.} For electrovacuum fluctuations this problem can be avoided since, without loss of generality, one can always work in the ``duality frame'' where the BH only carries electric charge (although the necessity of making such choice is clearly undesirable). However, this idea does not work in general if the Einstein--Maxwell theory is coupled to additional matter, i.e.~in the presence of sources $\bold{J}$, $S_{\mu\nu}$. Indeed, in that case electric-magnetic duality transformations may not even be defined in the first place, and even if they are, the ``duality frame'' where the BH is purely electric will in general contain magnetically charged currents, so $d\delta \mathcal{F}\ne0$ and $\delta \mathcal{A}_{\mu}$ does not exist. 

Here we introduce an alternative procedure that yields decoupled equations in all cases. The key observation is that one should work with variables that are manifestly invariant under electric-magnetic duality. Recalling that under duality transformations \eqref{phasetrafo} the background charge $\bold{C}$ behaves as $\bold{C}\mapsto e^{i\alpha}\bold{C}$, it is clear that $\bar{\bold{C}}\delta \bold{F}$ and $\bar{\bold{C}} \bold{J}$ are duality-invariant quantities. Their expansion in harmonics reads 
\begin{align}\notag
\bar{\bold{C}}\delta \bold{F}=&\frac{1}{2!}i\varphi(y)Y\varepsilon_{ab}dy^{a}\wedge dy^{b}+\frac{1}{2!}\Phi(y)Y\epsilon_{AB}dz^{A}\wedge dz^{B}\\
&+\Big(i\varphi_{a}(y)Z_{A}+\gamma_{a}(y)X_{A}\Big)dy^{a}\wedge dz^{A}\, ,\\ \notag \\
\bar{\bold{C}}\bold{J}=&\mathcal{J}_{a}(y)Ydy^{a}+\left(\mathcal{J}(y)Z_{A}-i\mathcal{V}(y)X_{A}\right)dz^{A}\, ,
\end{align}
where the factors of $i$ and the signs are merely conventional. The monopole mode ($l=0$) corresponds to inducing a small change in mass and dyonic charge. Here we shall focus on the dynamical modes and assume henceforth $l\geq1$. The dipole $l=1$ needs to be treated separately since the gravitational degree of freedom becomes non-dynamical, so we shall consider first the multipoles $l\geq2$ where both gravitational and electromagnetic degrees of freedom fluctuate. Proceeding as we did for the gravitational fluctuation, we expand the gauge-invariant quantity $\bar{\bold{C}} \delta \bold{F}+\pounds_{\eta} \bar{\bold{C}} \bold{F}$ in harmonics,\footnote{We notice that in the covariant language introduced in Section \ref{S1} the background field strength \eqref{RN} reads $\bold{F}=-\frac{i \bold{C}}{r^{2}}\ \frac{\varepsilon_{ab}}{2!}\ dy^{a}\wedge dy^{b}+\bold{C}\ \frac{\epsilon_{AB}}{2!}dz^{A}\wedge dz^{B}$.}
\begin{align}\notag
\bar{\bold{C}} \delta \bold{F}+\pounds_{\eta} \bar{\bold{C}} \bold{F}=&\frac{1}{2!}i\tilde{\varphi}(y)Y\varepsilon_{ab}dy^{a}\wedge dy^{b}\\  \label{gaugeINV}
&+\frac{1}{2!}\tilde{\Phi}(y)Y\epsilon_{AB}dz^{A}\wedge dz^{B}\\ \notag
&+\Big(i\tilde{\varphi}_{a}(y)Z_{A}+\tilde{\gamma}_{a}(y)X_{A}\Big)dy^{a}\wedge dz^{A}\, ,
\end{align}
where $\eta_{\mu}$ is given in \eqref{vec1}-\eqref{vec3}, and the explicit form of the various components $\tilde{\varphi}, \tilde{\Phi},...$ in terms of the original ones $\varphi, \Phi,...$ is given in Appendix \ref{SHandEE}. The linearised Maxwell equations \eqref{max1} and \eqref{max2} now read 
\begin{align}
&\varepsilon^{cd}\nabla_{c}\tilde{\gamma}_{d}=4\pi  \mathcal{J}\, ,\\
&\varepsilon^{cd}\nabla_{c}\tilde{\varphi}_{d}-\tilde{\varphi}=4\pi \mathcal{V}\, , \\
&\nabla_{a}\tilde{\Phi}-\lambda^{2}\tilde{\gamma}_{a}=-4\pi r^{2}\varepsilon_{ab}\mathcal{J}^{b}\, ,
\end{align}
and 
\begin{align}
&\tilde{\varphi}+\frac{\tilde{\Phi}}{r^{2}}=-\frac{ \bar{\bold{C}}\bold{C}}{2r^{2}}\left(\tilde{h}^{a}_{\ a}-\frac{2\tilde{k}}{r^{2}}\right)\, ,\\
&\tilde{\varphi}_{a}+\varepsilon_{ab}\tilde{\gamma}^{b}=-i\frac{\bar{\bold{C}}\bold{C}}{r^{2}}\tilde{j}_{a}\, .
\end{align}
Using these equations one immediately finds that the gauge-invariant components \eqref{TtildeCOMPS} associated to the Maxwell energy-momentum tensor are  
\begin{equation}\label{emfluct}
\begin{aligned}
\tilde{\theta}_{ab}&=-\frac{1}{r^{4}}\left[\tilde{\Phi}^{+}g_{ab}+\bar{\bold{C}}\bold{C}\left(\tilde{h}_{ab}-\frac{2}{r^{2}}\tilde{k}g_{ab}\right)\right] \, , \\  \\
\tilde{\theta}_{a}&=-\frac{1}{\lambda^{2}r^{2}}\left(\nabla_{a}\tilde{\Phi}^{+}+4\pi r^{2}\varepsilon_{a}^{\ b}\mathcal{J}^{+}_{b}\right) \, , \\  \\
\tilde{\theta}&=\frac{1}{r^{2}}\left[\tilde{\Phi}^{+}-\frac{\bar{\bold{C}}\bold{C}}{r^{2}}\tilde{k}\right] \, , \\  
\tilde{\rho}_{a}&=-\frac{i}{\lambda^{2}r^{2}}\left[\varepsilon_{ab}\nabla^{b}\tilde{\Phi}^{-}+4\pi r^{2} \mathcal{J}_{a}^{-}+i\lambda^{2}\frac{\bar{\bold{C}}\bold{C}}{r^{2}}\tilde{j}_{a}\right]\, ,
\end{aligned}
\end{equation}
while $\tilde{\rho}$ and $\tilde{\sigma}$ vanish in this theory. As we shall see immediately, the scalars $\tilde{\Phi}^{+}$ and $\tilde{\Phi}^{-}$ are the master variables of the generalised even and odd sectors of the Maxwell field. They are simply given by
\begin{equation}
\tilde{\Phi}^{\pm}\equiv\tilde{\Phi}\pm\tilde{\Phi}^{*}\, ,
\end{equation}
where the superscript $*$ means ``complex harmonic conjugation'', defined as $(A^{(l,m)})^{*}\equiv(-1)^{m}\bar{A}^{(l,-m)}$ where $A^{(l,m)}$ are the harmonic components of a generic tensor field $A$ (notice that if $A$ is a real tensor then $(A^{(l,m)})^{*}=A^{(l,m)}$, as follows from the property $\bar{Y}^{(l,m)}=(-1)^{m}Y^{(l,-m)}$ of spherical harmonics, but in general $(A^{(l,m)})^{*}\ne A^{(l,m)}$ if $A$ is complex). From equations \eqref{emfluct} it follows that the Maxwell field couples to the even and odd gravitational sectors only via $\tilde{\Phi}^{+}$ and $\tilde{\Phi}^{-}$, respectively. Furthermore, as can be readily verified $\tilde{\Phi}^{\pm}$ satisfy the second order equations 
\begin{widetext}
\begin{align}\label{MaxP}
\left(\square-\frac{\lambda^{2}}{r^{2}}\right)\tilde{\Phi}^{+}=&\lambda^{2}\frac{\bar{\bold{C}}\bold{C}}{r^{2}}\left(\tilde{h}^{c}_{\ c}-\frac{2}{r^{2}}\tilde{k}\right)-4\pi\lambda^{2}\mathcal{V}^{+} -4\pi \varepsilon^{ab}\nabla_{a}\left(r^{2}\mathcal{J}^{+}_{b}\right)\, , \\ \notag \\ \label{MaxM}
\left(\square-\frac{\lambda^{2}}{r^{2}}\right)\tilde{\Phi}^{-}=&-2i \lambda^{2}\bar{\bold{C}}\bold{C} \varepsilon^{ab}\nabla_{a}\left(\frac{\tilde{j}_{b}}{r^{2}}\right)-4\pi \lambda^{2}\mathcal{V}^{-} -4\pi \varepsilon^{ab}\nabla_{a}\left(r^{2}\mathcal{J}^{-}_{b}\right)\, , 
\end{align}
\end{widetext}
where $\square\equiv \nabla^{a}\nabla_{a}$ is the wave operator of $g_{ab}$, and again $\tilde{\Phi}^{+}$ and $\tilde{\Phi}^{-}$ couple only to the even and odd gravitational sectors, respectively.

We conclude that the equations decouple in two sets. One involves the the even gravitational variables and $\tilde{\Phi}^{+}$, and is governed by equations \eqref{even} and \eqref{MaxP}. Likewise, the other set involves the odd gravitational variables and $\tilde{\Phi}^{-}$, and is subject to equations \eqref{odd} and \eqref{MaxM}. The source terms appearing in these equations are completely general, and it is only assumed that they satisfy the conservation laws \eqref{consC1} and \eqref{consC2}, which are required by consistency. The same analysis holds for the dipole modes $l=1$ (setting to zero the appropriate components, see Appendix \ref{SHandEE}), bearing in mind that now the tilded variables are in general not gauge-invariant (but the equations are, of course). These two sets of equations generalise the usual even and odd sectors of the fluctuations of a purely electric BH, coupled to purely electric sources (see e.g. \cite{Zerilli:1974ai,PhysRevD.10.1057,Chandrasekhar:1985kt,Kodama:2003kk}). We expect that a similar procedure can be applied to decouple the fluctuations of spherically symmetric BHs in more general theories that exhibit some notion of self-duality, such as the BHs in axion-dilation gravity \cite{Kallosh:1993yg,Mitsios:2021zrn}. However, including BH rotation most likely requires following an approach based on the Newman--Penrose formalism \cite{Newman:1961qr}, even though a complete decoupling is not expected to take place in that case \cite{Dias:2015wqa}. 

Before closing this section it is instructive to write $\tilde{\Phi}^{\pm}$ in terms of the real Maxwell field strength, $\delta \mathcal{F}$, whose harmonic expansion can be written as 
\begin{align}\notag
\delta \mathcal{F}=&\frac{1}{2!}\mathcal{E}(y)Y\varepsilon_{ab}dy^{a}\wedge dy^{b}+\frac{1}{2!}\mathcal{B}(y)Y\epsilon_{AB}dz^{A}\wedge dz^{B}\\
&+\Big(\mathcal{E}_{a}(y)Z_{A}+\mathcal{B}_{a}(y)X_{A}\Big)dy^{a}\wedge dz^{A}\, .
\end{align}
Plugging this into the definition of $\tilde{\Phi}^{\pm}$ given in \eqref{gaugeINV}, one finds \footnote{Assuming for simplicity that $\delta \mathcal{F}$ is in the gauge $\eta_{\mu}[h]=0$, which always exists.}
\begin{align}\label{comb1}
\tilde{\Phi}^{+}&=2 P \mathcal{B}-2 Q \left[r^{2}\mathcal{E}+Q\left(\frac{\tilde{h}^{a}_{\ a}}{2}-\frac{\tilde{k}}{r^{2}}\right)\right]\, ,\\ \label{comb2}
\tilde{\Phi}^{-}&=2 i Q \mathcal{B}+2i P \left[r^{2}\mathcal{E}+Q\left(\frac{\tilde{h}^{a}_{\ a}}{2}-\frac{\tilde{k}}{r^{2}}\right)\right]\, .
\end{align}
Taking a background BH which is purely electric ($P=0$ and $Q\ne0$), we see that it is only the electric field $\mathcal{E}$ that couples to the even gravitational sector, while only the magnetic field $\mathcal{B}$ couples to the odd one, a fact that was first observed in \cite{Zerilli:1974ai}. However, this works the other way around on a purely magnetic BH ($Q=0$ and $P\ne0$), while we find that in the most general case ($Q\ne0$ and $P\ne0$) it is precisely the combinations \eqref{comb1} and \eqref{comb2} that couple to the even and odd gravitational sectors, respectively.

\subsection{Master Wave Equations}\label{S3}
Having decoupled the Einstein--Maxwell equations into our generalised even and odd sectors, we are in conditions of deriving master wave equations governing the dynamics of each sector. Such derivation is straightforward and is inspired by previous works in the literature (e.g.~\cite{Zerilli:1974ai,PhysRevD.10.1057,Chandrasekhar:1985kt,Martel:2005ir,Kodama:2003kk,Ishibashi:2011ws}), although the approach here is manifestly covariant. In this section we only report the final equations and source terms, and leave the details of the derivation to Appendix \ref{MasterWaves}.

Each sector is governed by two decoupled wave equations, corresponding to the gravitational and electromagnetic modes. These can be cast in the form 
\begin{equation}\label{masteq}
\left(\square-V_{1,2}^{\pm}\right)\Psi_{1,2}^{\pm}=\bold{S}_{1,2}^{\pm}\, ,
\end{equation}
where $+,-$ refer to the generalised even and odd sectors, and $1,2$ refer to the electromagnetic and the gravitational modes, respectively. The wave operator $\square=\nabla^{a}\nabla_{a}$ is associated to the two-dimensional Lorentzian background metric $g_{ab}$, and the potentials can be written in the compact form  
\begin{equation}\label{pots}
V^{\pm}_{1,2}=\pm q_{2,1}\frac{\text{d}}{\text{d}r}W_{1,2}+q_{2,1}^{2}f^{-1}W^{2}_{1,2}+\lambda^{2}\left(\lambda^{2}-2\right)f^{-1}W_{1,2}
\end{equation}
in terms of the gravitational and electromagnetic ``super potentials'',
\begin{equation}
W_{1,2}(r)=\frac{f(r)}{r\left((\lambda^{2}-2)r+q_{2,1}\right)}\, ,
\end{equation}
where we have introduced the constants $ q_{1}=3M+\Delta$, $q_{2}=3M-\Delta$, and $\Delta=\sqrt{9 M^{2}+4\bar{\bold{C}}\bold{C}(\lambda^{2}-2)}$. Equation \eqref{pots} generalises to dyonic RN(A)dS BHs the relation found by Chandrasekar between the even and odd fluctuations of an electric RN BH \cite{Chandrasekhar:1985kt}. The master variables $\Psi_{1,2}^{\pm}$ are related to the field variables introduced in the previous sections by
\begin{widetext}
\begin{align}
\left( \begin{array}{@{}c@{}} \Psi^{-}_{1} \\ \Psi^{-}_{2} \end{array} \right)&=\frac{1}{2\Delta \mathcal{A}^{-}}\left(\begin{array}{@{}cc@{}} -q_{2} & -\frac{i(\lambda^{2}-2)}{\lambda^{2}}  \\  \frac{q_{1}}{\alpha^{-}} &  \frac{i(\lambda^{2}-2)}{\lambda^{2}\alpha^{-}} \\ \end{array}\right)\left( \begin{array}{@{}c@{}} -\frac{r^{3}}{2}\varepsilon^{ab}\nabla_{a}\left(\frac{\tilde{j}_{b}}{r^{2}}\right)+\frac{i}{\lambda^{2}r}\tilde{\Phi}^{-} \\  \tilde{\Phi}^{-} \end{array} \right)\\ \notag \\
\left( \begin{array}{@{}c@{}} \Psi^{+}_{1} \\ \Psi^{+}_{2} \end{array} \right)&=\frac{1}{2\Delta\mathcal{A}^{+}}\left(\begin{array}{@{}cc@{}} -q_{2} & \frac{2}{\lambda^{2}}  \\ \frac{q_{1}}{\alpha^{+}} &  -\frac{2}{\lambda^{2}\alpha^{+}}  \\ \end{array}\right)\left( \begin{array}{@{}c@{}} \left(t^{a}r^{b}p_{ab}-t^{a}\nabla_{a}(\tilde{k}/r)\right)/U(r) \\  t^{a}\nabla_{a}\tilde{\Phi}^{+}+2\lambda^{2}\frac{\bar{\bold{C}}\bold{C}}{r}\left[\left(t^{a}r^{b}p_{ab}-t^{a}\nabla_{a}(\tilde{k}/r)\right)/U(r)\right] \end{array} \right)
\end{align}
\end{widetext}
where $\mathcal{A}^{\pm}$ and $\alpha^{\pm}$ are arbitrary non-zero constants, $U(r)=\frac{6M+r(\lambda^{2}-2)}{r}-4\frac{\bar{\bold{C}}\bold{C}}{r^{2}}$, and $p_{ab}=\tilde{h}_{ab}-(1/2)\tilde{h}^{c}_{\ c}g_{ab}$ is the traceless part of the gauge-invariant metric fluctuation $\tilde{h}_{ab}$. 

Before discussing the source terms $\bold{S}_{1,2}^{\pm}$, it is worth stressing here that our equations \eqref{masteq} are formally similar to those obtained in the literature for purely electric RN BHs, \cite{Zerilli:1974ai,Chandrasekhar:1985kt,Kodama:2003kk}. In particular, the potentials in \eqref{pots} are the same as those obtained in \cite{Kodama:2003kk} just replacing $ \bar{\bold{C}}\bold{C}\to Q^{2}$, as one would expect form electric-magnetic duality. There are, however, two important differences. First, our equations are written for variables which are manifestly gauge- and duality-invariant, and are defined irrespective of the charge configuration of the background BH. Second, and most important, the source terms in our expressions are new and generalise those in \cite{Zerilli:1974ai,Chandrasekhar:1985kt,Kodama:2003kk} by including, in a covariant and duality-invariant guise, the coupling of the most general dyonic matter sources to a fluctuation of a dyonic RN(A)dS BH \eqref{RN}. Explicitly, $\bold{S}_{1,2}^{\pm}$ are 
\begin{align}
\bold{S}_{1}^{-}&=\frac{-1}{2\mathcal{A}^{-}\Delta}\left(q_{2} S_{\mathfrak{g}}^{-}+i\frac{\lambda^{2}-2}{\lambda^{2}}S_{\mathfrak{a}}^{-}\right)\, , \\
\bold{S}_{2}^{-}&=\frac{1}{2\mathcal{A}^{-}\Delta\alpha^{-}}\left(q_{1}S_{\mathfrak{g}}^{-}+i\frac{\lambda^{2}-2}{\lambda^{2}}S_{\mathfrak{a}}^{-}\right)\, , \\
\bold{S}_{1}^{+}&=\frac{-1}{2  \mathcal{A}^{+} \Delta}\left(q_{2}S^{+}_{\mathfrak{g}}-\frac{2}{\lambda^{2}}S^{+}_{\mathfrak{a}}\right)\, ,\\
\bold{S}_{2}^{+}&=\frac{1}{2  \mathcal{A}^{+} \Delta \alpha^{+}}\left(q_{1}S_{\mathfrak{g}}^{+}-\frac{2}{\lambda^{2}}S^{+}_{\mathfrak{a}}\right)\, ,
\end{align}
where $S_{\mathfrak{a},\mathfrak{g}}^{\pm}$ are given by the covariant expressions
\begin{widetext}
\begin{align}\notag
S^{+}_{\mathfrak{g}}=&S^{(1)}_{\mathfrak{g}}\ t^{c}\nabla_{c}\left(r^{a}r^{b}\Sigma_{ab}\right)+S^{(2)}_{\mathfrak{g}}\ r^{c}\nabla_{c}\left(t^{a}r^{b}\Sigma_{ab}\right) +S^{(3)}_{\mathfrak{g}}\ t^{a}r^{b}\Sigma_{ab}+S^{(4)}_{\mathfrak{g}}\ t^{c}\nabla_{c}\left(r^{a}\Sigma_{a}\right)+S^{(5)}_{\mathfrak{g}} r^{c}\nabla_{c}\left(t^{a}\Sigma_{a}\right)\\  \label{SI}
&+S^{(6)}_{\mathfrak{g}}\ t^{a}\Sigma_{a} +S^{(7)}_{\mathfrak{g}}\ t^{c}\nabla_{c}\left(t^{a}\mathcal{J}^{+}_{a}\right)+S^{(8)}_{\mathfrak{g}}\ r^{c}\nabla_{c}\left(r^{a}\mathcal{J}^{+}_{a}\right)+S^{(9)}_{\mathfrak{g}}\ r^{a}\mathcal{J}^{+}_{a}+S^{(10)}_{\mathfrak{g}}\ t^{a}\nabla_{a}\mathcal{S}\, , \\ \notag \\ \notag
S^{+}_{\mathfrak{a}}=&2\lambda^{2}\frac{\bar{\bold{C}}\bold{C}}{r}S_{\mathfrak{g}}^{+}+S^{(1)}_{\mathfrak{a}}t^{a}r^{b}\Sigma_{ab}+S^{(2)}_{\mathfrak{a}}t^{a}\Sigma_{a}+S^{(3)}_{\mathfrak{a}}t^{a}\nabla_{a}\mathcal{S}+S^{(4)}_{\mathfrak{a}}\ r^{a}\mathcal{J}^{+}_{a}+S^{(5)}_{\mathfrak{a}}\ t^{c}\nabla_{c}\left[\varepsilon^{ab}\nabla_{a}\left(r^{2}\mathcal{J}_{b}^{+}\right)\right]\\
&+S^{(6)}_{\mathfrak{a}}\ t^{a}\nabla_{a}\mathcal{V}^{+}\, , \\ \notag \\ 
S^{-}_{ \mathfrak{g}}=&r\varepsilon^{ab}\left(\nabla_{a}\Upsilon_{b}-\frac{4\pi i}{\lambda^{2}}\nabla_{a}\mathcal{J}_{b}^{-}\right)\, ,\\ \notag \\ \label{SF}
S^{-}_{\mathfrak{a}}=&-4\pi\left(\lambda^{2}\mathcal{V}^{-}+\varepsilon^{ab}\nabla_{a}\left(r^{2}\mathcal{J}^{-}_{b}\right)\right)\, , 
\end{align}
\end{widetext}
with $S_{\mathfrak{a},\mathfrak{g}}^{(i)}$ some functions of $r$ reported in equations \eqref{Sa}-\eqref{Sg} of Appendix \ref{MasterWaves}. 

If the background BH and the currents are purely electric, our source terms reduce to those in \cite{Zerilli:1974ai,Chandrasekhar:1985kt,Kodama:2003kk}, as expected. However, for a general charge configuration there are additional terms that excite new channels of gravitational and electromagnetic radiation. To see this, consider a typical source term such as $\mathcal{J}^{-}_{a}$ (see \eqref{SI}-\eqref{SF}), and assume that the background BH is purely electric, so $\bold{C}=-iQ$. Then $\mathcal{J}^{-}_{a}$ reads
\begin{align}
\mathcal{J}^{-}_{a}&=2 i Q\int d\Omega \ \bar{Y} J_{a}^{(m)}\, ,
\end{align}
where $J_{a}^{(m)}$ is the real magnetic current in \eqref{Mm1}. Thus, if the currents are purely electric too, i.e. $J_{a}^{(m)}=0$, the terms associated to $\mathcal{J}^{-}_{a}$ drop from the sources $\eqref{SI}$-$\eqref{SF}$ and one recovers the known results for purely electric BHs and currents \cite{Zerilli:1974ai,Chandrasekhar:1985kt,Kodama:2003kk}. In the presence of magnetic currents, though, $J_{a}^{(m)}\ne0$ and there are additional contributions to the gravitational and electromagnetic radiation of intensity $\sim Q J^{(m)}$ (or $\sim P J^{(e)}$ in the case of a magnetic background BH with charge $P$ and an electric current $J^{(e)}$). As discussed in \cite{Dyson:2023ujk}, these new modes that emerge from an ``electric-magnetic'' interaction (as opposed to an ``electric-electric'' or ``magnetic-magnetic'' one) exhibit a rich and interesting phenomenology that has no counterpart in purely electric configurations.

To conclude, we remark that the dipolar modes $l=1$ are entirely governed by the electromagnetic degree of freedom alone, both in the even and the odd sectors. A detailed treatment of these modes is provided in Appendix \ref{MasterWaves}.

\section{Discussion}\label{Disc}
We extended the usual harmonic description of perturbations of spherically symmetric BHs in the Einstein--Maxwell theory, and made electric-magnetic self-duality manifest. This allowed us to deal with the coupling of the traditional even and odd fluctuations that takes place whenever the BH is magnetically charged. Our generalised even and odd sectors, which are manifestly gauge- and duality-invariant, satisfy decoupled equations for arbitrary values of the electric and magnetic charges of both the BH and the sources. Furthermore, the dynamics of the electromagnetic field is encoded in a single, complex scalar $\tilde{\Phi}$ whose independent components $\tilde{\Phi}^{+}$ and $\tilde{\Phi}^{-}$ describe the even and odd fluctuations, respectively. We have also provided decoupled master wave equations that govern each sector and include the most general dyonic source terms in a manifestly covariant and duality-invariant form.

Our results are important in the context of GW physics because they lead to robust theoretical predictions about GW generation and electromagnetic radiation by the most general spherically symmetric BHs of the Einstein--Maxwell theory. In particular, our wave equations are not restricted to the Newtonian regime \cite{Liu:2020cds,Liu:2020vsy,Liu:2020bag,Liu:2022wtq,Chen:2022qvg} and include both strong field and relativistic effects in the extreme mass ratio limit, thus being relevant for low frequency GW detectors such as LISA. In addition, the fact that they are valid for the most general charge configuration allows an exploration of regions of parameter space that include phenomena with no counterpart in purely electric set ups \cite{Gupta:2021rod,Zerilli:1974ai,Johnston:1974vf,Bozzola:2020mjx,Zilhao:2012gp,Zilhao:2013nda,Liebling:2016orx}, such as those exhibited during accretion of electrically charged matter by magnetic BHs \cite{HGed,Garfinkle:1990zx,Dyson:2023ujk}. This could lead to novel ways of constraining the parameters of dyonic RN BHs from observations. Finally, from the perspective of beyond-the-Standard-Model physics and some dark matter models~\cite{DeRujula:1989fe,Perl:1997nd,Holdom:1985ag,Sigurdson:2004zp,Davidson:2000hf,McDermott:2010pa,Cardoso:2016olt,Khalil:2018aaj,Bai:2019zcd,Gupta:2021rod,Kritos:2021nsf}, our results can help elucidating signatures of new physics.

\vspace*{0.5cm}
\begin{acknowledgments}
We thank Vitor Cardoso, Gregorio Carullo, Juan Garc\'ia-Bellido, Akihiro Ishibashi, Caio Macedo and Tom\'as Ort\'in for stimulating and fruitful conversations. We also thank Vitor Cardoso for suggestions in previous versions of this paper.
This work was supported by VILLUM Foundation (grant no. VIL37766) and the DNRF Chair program (grant no. DNRF162) by the Danish National Research Foundation. %
\end{acknowledgments}
\bibliography{Bibliography}



\onecolumngrid

\appendix

\section{Spherical Harmonics and Linearised Einstein Equations}\label{SHandEE}
\subsection{Spherical Harmonics}
The spherical harmonics in the 2-sphere are defined by the eigenvalue equation
\begin{equation}
\left(D^{A}D_{A}+\lambda^{2}\right)Y^{(l,m)}=0\, , 
\end{equation}
where $\lambda^{2}=l(l+1)$ and $l=0,1,2 ...$ with $m\in \mathbb{Z}$ and $\lvert m\rvert \leq l$. We choose to normalise them so that the following orthonormality condition holds,
\begin{equation}
\begin{aligned}
\int d\Omega \  \bar{Y}^{(l',m')}Y^{(l,m)} =\delta_{l'l}\delta_{m'm}\, ,
\end{aligned}
\end{equation}
where $d\Omega$ is the volume element on the 2-sphere. The even tensor harmonics are (dropping the $(l,m)$ superscript) 
\begin{equation}
\begin{aligned}
Z_{A}\equiv D_{A}Y\, ,\ \ \ U_{AB}\equiv \Omega_{AB} Y\, , \ \ \ V_{AB}\equiv D_{A}D_{B}Y+\frac{\lambda^{2}}{2}U_{AB}\, ,
\end{aligned}
\end{equation}
while the odd tensor harmonics read
\begin{align}
X_{A}\equiv\epsilon_{A}^{\ B}Z_{B}\, ,\ \ W_{AB}\equiv D_{(A}X_{B)}\, .
\end{align}
The orthogonality properties of tensor harmonics follow straightforwardly from their definitions. As indicated in the main text, the lower order harmonics $l=0,1$ are special. For $l=0$, $Y$ is a constant, so $Z_{A}=X_{A}=W_{AB}=0$, while $V_{AB}$ is not defined (or equivalently, it is proportional to $U_{AB}$). In particular, this means that only even harmonics exist for the monopole $l=0$. For the dipole $l=1$, $X_{A}$ is a Killing vector on the sphere, so $W_{AB}=V_{AB}=0$.
\subsection{Gauge-invariant Variables}
In the basis of harmonics introduced above, the components of the gauge-invariant variables associated to the metric and energy-momentum fluctuation, introduced in equations \eqref{htildeCOMPS} and \eqref{TtildeCOMPS}, read 
\begin{align}
\tilde{h}_{ab}&=h_{ab}+2\nabla_{(a}\eta_{b)}\, ,\\
\tilde{k}&=k+2r r^{a}\eta_{a}-\lambda^{2}\eta\, , \\
\tilde{j}_{a}&=j_{a}+r^{2}\nabla_{a}\left(\frac{\upsilon}{r^{2}}\right)\, 
\end{align}
and
\begin{align}
\tilde{\theta}_{ab}&=\theta_{ab}+\eta^{c}\nabla_{c}T_{ab}+T_{cb}\nabla_{a}\eta^{c}+T_{ac}\nabla_{b}\eta^{c}\,, \\
\tilde{\theta}_{a}&=\theta_{a}+T_{ab}\eta^{b}+r^{2}\mathcal{T}\nabla_{a}\left(\frac{\eta}{r^{2}}\right)\,, \\
\tilde{\theta}&=\theta +\eta^{d}\nabla_{d}\left(r^{2}\mathcal{T}\right)-\lambda^{2}P\eta\,, \\
\tilde{\sigma}&=\sigma+2\mathcal{T}\eta \,, \\
\tilde{\rho}_{a}&=\rho_{a}+r^{2}\mathcal{T}\nabla_{a}\left(\frac{\upsilon}{r^{2}}\right) \,, \\
\tilde{\rho}&= \rho+2\mathcal{T} \upsilon\,.
\end{align}
As mentioned in the main text, these variables are gauge-invariant for the harmonic modes with $l\geq2$. Let us now comment on the lower multipoles $l=0,1$. For either of these modes there is no analogue of the perturbation dependent vector field $\eta_{\mu}[h]$ that can be used to compensate gauge transformations and construct gauge-invariant variables. The variables we will work with in those cases are defined just as the tilded variables above, but setting to zero the appropriate components of $\eta_{\mu}$ ($\eta_{\mu}=0$ for $l=0$, and $\chi=\upsilon=0$ for $l=1$). While it is possible to write covariant equations in terms of those variables, one should bear in mind they are not gauge-invariant in general. One important exception in the odd sector with $l=1$ is the combination
\begin{equation}
    \tau_{a}\equiv\tilde{\rho}_{a}-\mathcal{T} \tilde{j}_{a}\, ,
\end{equation}
as well as the exterior derivatives $\nabla_{[a}(\tilde{j}_{b]}/r^{2})$ and $\nabla_{[a}(\tilde{\rho}_{b]}/r^{2}\mathcal{T})$. Finally, we also report here the gauge-invariant Maxwell variables introduced in equation \eqref{gaugeINV}, which explicitly read
\begin{align}
\tilde{\varphi}=&\varphi-  \bar{\bold{C}}\bold{C}\nabla_{a}\left(\frac{\eta^{a}}{r^{2}}\right)\, ,\\
\tilde{\varphi}_{a}=&\varphi_{a}-i \bar{\bold{C}}\bold{C}\left(\nabla_{a}\left(\frac{\upsilon}{r^{2}}\right)-i\varepsilon_{ab}\frac{\eta^{b}}{r^{2}}\right)\, , \\
\tilde{\gamma}_{a}=&\gamma_{a}- \bar{\bold{C}}\bold{C} \nabla_{a}\left(\frac{\eta}{r^{2}}\right)\, , \\
\tilde{\Phi}=&\Phi- \frac{\bar{\bold{C}}\bold{C}}{r^{2}}\lambda^{2}\eta\, .
\end{align}
\subsection{Linearised Einstein Equations}
The linearised Einstein's equations are
\begin{equation}\label{EinEq}
\delta \hat{G}_{\mu\nu}+\Lambda h_{\mu\nu}-\delta T_{\mu\nu}=S_{\mu\nu}\, ,
\end{equation}
where $\delta T_{\mu\nu}$ is the variation of the energy-momentum tensor associated to the fields whose background value is non-zero, and $S_{\mu\nu}$ is a first order source. We shall denote the left hand side of \eqref{EinEq} by
\begin{equation}
\mathcal{G}_{\mu\nu}\equiv \delta \hat{G}_{\mu\nu}+\Lambda h_{\mu\nu}-\delta T_{\mu\nu}\, .
\end{equation}
Before considering the four-dimensional case, it is instructive to write explicit expressions for $\mathcal{G}_{\mu\nu}$ in the general spacetime with structure $\mathcal{N}^{N}\times \mathbb{S}^{2}$ described in Section \ref{S1}. One finds 
\begin{align}\notag
 \mathcal{G}_{ab}=&\Biggl \{ G^{(1)}_{ab}[\tilde{h}]-\tilde{\theta}_{ab}\\ \notag
 &-\frac{1}{r}\left[\nabla_{a}\nabla_{b}\left(\frac{\tilde{k}}{r}\right)-\square \left(\frac{\tilde{k}}{r}\right) g_{ab}\right] + \frac{r^{c}}{r}\nabla_{c}\left(\frac{\tilde{k}}{r^{2}}\right)g_{ab}+\left[\nabla_{a}r_{b}-\left(\nabla_{c}r^{c}+\frac{\lambda^{2}-2}{2r}\right)g_{ab}\right]\frac{\tilde{k}}{r^{3}}\\ \notag 
 &+\frac{r^{c}}{r}\left(K_{cab}-K_{c}g_{ab}\right)+\frac{\lambda^{2}}{2 r^{2}}\left(\tilde{h}_{ab}-\tilde{h}^{c}_{\ c}g_{ab}\right)+\left(2\frac{\nabla_{c}r^{c}}{r}+\frac{H}{r^{2}}+\Lambda \right)\tilde{h}_{ab}-\left(\frac{r_{c}r_{d}}{r^{2}}+\frac{2}{r} \nabla_{c}r_{d}\right)\tilde{h}^{cd} g_{ab}\Biggr \}Y\\ \label{Eab}
 \equiv& E_{ab}Y \, , \\ \notag \\ \notag
\mathcal{G}_{aA}=& \frac{1}{2}\Biggl\{  \frac{1}{2}K_{a}-\frac{r^{2}}{2}\nabla_{a}\left(\frac{\tilde{h}^{b}_{\ b}}{r^{2}}\right)-\nabla_{a}\left(\frac{\tilde{k}}{r^{2}}\right)-2\tilde{\theta}_{a}\Biggr\}Z_{A}\\ \notag
&+\left\{r^{-2}\nabla^{b}\left(r^{4}\nabla_{[a}v_{b]}\right)+\frac{\lambda^{2}-2}{2}v_{a}+r^{2}P v_{a} -\tilde{\rho}_{a}\right\}X_{A}\\
\equiv& E_{a}Z_{A}+O_{a}X_{A} \, ,\\ \notag \\ \notag
\mathcal{G}_{AB}=&\left(\nabla_{a}\left(r^{2}v^{a}\right)-\tilde{\rho}\right)W_{AB}-\left(\frac{1}{2}\tilde{h}^{a}_{\ a}+\tilde{\sigma} \right)V_{AB}\\ \notag
&+\frac{1}{2}\Biggl\{r\square\left(\frac{\tilde{k}}{r}\right)+\left(2\mathcal{T}(r)-\frac{\nabla_{a}r^{a}}{r}\right)\tilde{k}-\frac{1}{2}\nabla_{a}\left(r^{2}K^{a}\right)+r^{2}\left(R^{ab}-2\frac{\nabla^{a}r^{b}}{r}\right)\tilde{h}_{ab}\\ \notag
&+\frac{r^{2}}{2}\left(\square-\frac{\lambda^{2}}{r^{2}}\right)\tilde{h}^{a}_{\ a}-2\tilde{\theta}\Biggr\}U_{AB} \\ \label{Eps}
\equiv&O W_{AB}+ \mathcal{E} V_{AB}+E U_{AB} \, ,
\end{align}
where we are implicitly assuming the sum over harmonics (i.e. we omit writing the harmonic indices $(l,m)$ and the sums $\sum_{l,m}$), and we introduced 
\begin{align}
K_{abc}&\equiv\nabla_{b}\tilde{h}_{ca}+\nabla_{c}\tilde{h}_{ba}-\nabla_{a}\tilde{h}_{bc}\, , \\
K_{a}&\equiv\tensor{K}{_a^b_b}\, , \\
v_{a}&\equiv r^{-2}\tilde{j}_{a}\, .
\end{align}
In addition, $R^{(1)}_{ab}[\tilde{h}]$ and $G^{(1)}_{ab}[\tilde{h}]$ denote the linear operators that result by formally expanding the Ricci and Einstein tensor of $g_{ab}$ to first order in a fluctuation $\tilde{h}_{ab}$, that is,
\begin{align}
R^{(1)}_{ab}[\tilde{h}]=&\frac{1}{2}\left(\Delta_{L}\tilde{h}_{ab}+\nabla_{(a}K_{b)}\right)+R_{c(a}\tilde{h}_{b)}^{\ c}\, , \\ \notag \\ \notag
G^{(1)}_{ab}[\tilde{h}]=&R^{(1)}_{ab}[\tilde{h}]-\frac{1}{2}R \tilde{h}_{ab}-\frac{1}{2}\left(g^{cd}R^{(1)}_{cd}[\tilde{h}]-\tilde{h}^{cd}R_{cd}\right)g_{ab}\, ,
\end{align}
and the usual Lichnerowicz operator reads $\Delta_{L}\tilde{h}_{ab}=-\square \tilde{h}_{ab}-2 R_{acbd}\tilde{h}^{cd}$. Equations \eqref{Eab}-\eqref{Eps} provide the left hand sides of equations \eqref{even} and \eqref{odd} in the main text, for the general spacetime $\mathcal{N}^{N}\times \mathbb{S}^{2}$. Before restricting to four dimensions, it is useful to expand in harmonics the conservation law of the (total) energy-momentum tensor, 
\begin{equation}
\delta\left(\hat{\nabla}^{\mu}T_{\mu\nu}\right)+\hat{\nabla}^{\mu}S_{\mu\nu}=0\, .
\end{equation}
The even parts of this equation are,
\begin{align} \notag
r^{-2}\nabla^{d}\left(r^{2}\Sigma_{da}\right)-\frac{\lambda^{2}}{r^{2}}\Sigma_{a}-2\frac{r_{a}}{r^{3}}\Sigma=&-r^{-2}\nabla^{d}\left[r^{2}\left(\tilde{\theta}_{da}-\tilde{h}^{b}_{\ d}T_{ba}\right)\right]+\frac{\lambda^{2}}{r^{2}}\tilde{\theta}_{a}+2\frac{r_{a}}{r^{3}}\tilde{\theta}\\ \notag
&-\frac{1}{2}\left[\nabla_{d}\left(\tilde{h}^{f}_{\ f}+2\frac{\tilde{k}}{r^{2}}\right) \right]T^{d}_{\ a}+\frac{1}{2}T^{df}\nabla_{a}\tilde{h}_{df}+P\left(\nabla_{a}\left(\frac{\tilde{k}}{r^{2}}\right)-2\frac{r_{a}}{r^{3}}\tilde{k}\right)\, , \\ \notag \\ \notag
r^{-2}\nabla_{a}\left(r^{2}\Sigma^{a}\right)+\frac{\Sigma}{r^{2}}+\frac{2-\lambda^{2}}{2 r^{2}}\mathcal{S}=&-r^{-2}\nabla_{a}\left(r^{2}\tilde{\theta}^{a}\right)-\frac{\tilde{\theta}}{r^{2}}+\frac{1}{2}T_{ab}\tilde{h}^{ab}-\frac{P}{2}\left(\tilde{h}^{f}_{\ f}-2\frac{\tilde{k}}{r^{2}}\right)-\frac{2-\lambda^{2}}{2 r^{2}}\tilde{\sigma}\, ,
\end{align}
while the odd one is
\begin{align}\notag
\nabla_{a}\left(r^{2}\Upsilon^{a}\right)+\frac{(2-\lambda^{2})}{2}\Upsilon=&-\nabla_{a}\left[r^{2}\left(\tilde{\rho}^{a}-r^{2}Pv^{a}\right)\right] -\frac{(2-\lambda^{2})}{2}\tilde{\rho}\, .
\end{align}
In the four-dimensional case, where the spacetime has structure $\mathcal{N}^{2}\times \mathbb{S}^{2}$, some remarkable simplifications take place. First, as is well known from the Gauss--Bonnet theorem, the Einstein tensor of any 2-dimensional (pseudo)-Riemannian manifold vanishes identically, $G_{ab}=0$ (not to be confused with $\hat{G}_{ab}$), so in particular $G^{(1)}_{ab}[\tilde{h}]=0$ for any $\tilde{h}_{ab}$. This removes from Einstein's equations all second order derivatives of
\begin{equation}
p_{ab}\equiv \tilde{h}_{ab}-(1/2)\tilde{h}^{c}_{\ c}g_{ab}\, ,
\end{equation}
the traceless part of the gauge-invariant metric fluctuation. Finally, the background matter and vacuum contributions to the equations of motion can be neatly separated by using the background equations 
\begin{align}
\frac{2}{r}\nabla_{a}r_{b}&=\left(\frac{R}{2}+\mathcal{T}-\Lambda\right)g_{ab}-\check{T}_{ab}\, , \\
\frac{r^{c}r_{c}-1}{r^{2}}&=\frac{1}{2}\left(T^{a}_{\ a}-R-2\mathcal{T}\right)\, ,
\end{align}
where $\check{T}_{ab}\equiv T_{ab}-(1/2)T^{c}_{\ c}g_{ab}$ is the traceless part of $T_{ab}$, to cast the even pieces $E_{ab}$ and $E$ of \eqref{Eab}-\eqref{Eps} in the form 
\begin{align}\notag
 E_{ab}=&-\frac{1}{r}\left[\nabla_{a}\nabla_{b}\left(\frac{\tilde{k}}{r}\right)-\square \left(\frac{\tilde{k}}{r}\right) g_{ab}\right] -\left[\frac{r^{c}r^{d}p_{cd}}{r^{2}}-\frac{r^{c}}{r}\nabla_{c}\left(\frac{\tilde{k}}{r^{2}}\right)\right]g_{ab}+\frac{r^{c}}{r}\left(K_{cab}-K_{c}g_{ab}\right)\\ \notag
 &-\frac{1}{2}\left[\check{T}_{ab}+\left(\frac{R}{2}+\mathcal{T}-\Lambda+\frac{\lambda^{2}-2}{r^{2}}\right)g_{ab}\right]\frac{\tilde{k}}{r^{2}} +\frac{1}{2}\left(R+2\mathcal{T}-2\Lambda+T^{a}_{\ a}+\frac{\lambda^{2}}{r^{2}}\right)p_{ab}\\ 
 &+\check{T}_{cd}p^{cd}g_{ab}+\frac{2\Lambda r^{2}-\left(\lambda^{2}+2\right)}{4 r^{2}}\tilde{h}^{f}_{\ f}g_{ab} \, , \\ \notag \\
 E=& \frac{1}{2}\left[ r\square\left(\frac{\tilde{k}}{r}\right)+\left(\mathcal{T}+\Lambda-\frac{R}{2}\right)\tilde{k} -\frac{1}{2}\nabla_{a}\left(r^{2}K^{a}\right)+r^{2}\check{T}_{ab}p^{ab}+\frac{r^{2}}{2}\left(\square-\frac{\lambda^{2}}{r^{2}}+2\left(\Lambda-\mathcal{T}\right)\right)\tilde{h}^{f}_{\ f}\right] \, .
\end{align}

\section{Derivation of Master Wave Equations}\label{MasterWaves}

In this appendix we provide the details of the derivation of the master wave equations \eqref{masteq}. We shall treat generalised even and odd sectors separately.

\subsection{Generalised Odd Sector}

In the set up of Section \ref{S1} for generic matter, the odd sector is governed by three equations, two coming from Einstein's equations and one from conservation of the energy-momentum tensor. They read, respectively, %
\begin{align}
&r^{-2}\nabla^{b}\left(r^{4}\nabla_{[a}v_{b]}\right)+\frac{(\lambda^{2}-2)}{2}v_{a}+r^{2}\mathcal{T} v_{a} -\tilde{\rho}_{a}=\Upsilon_{a}\, ,\\ \notag \\
&\nabla_{a}\left(r^{2}v^{a}\right)-\tilde{\rho}=\Upsilon\, , \\ \notag \\ \notag
&\nabla_{a}\left[r^{2}\left(\tilde{\rho}^{a}-r^{2}\mathcal{T}v^{a}\right)\right]+\frac{(2-\lambda^{2})}{2}\tilde{\rho}=-\nabla_{a}\left(r^{2}\Upsilon^{a}\right)-\frac{(2-\lambda^{2})}{2}\Upsilon\, .
\end{align}
Now, restricting to the four-dimensional set up, so our harmonic components are tensors on a 2-dimensional manifold $\mathcal{N}^{2}$ (see Section \ref{S1}), one can effectively reduce the order of the derivatives in the gravitational equation by one, using that $\nabla_{[a}v_{b]}$ must have the form
\begin{equation}
\nabla_{[a}v_{b]}=r^{-4}\Omega\varepsilon_{ab} \, ,
\end{equation}
for some function $\Omega$ (explicitly, $\Omega=-(r^{4}/2)\varepsilon^{ab}\nabla_{a}v_{b}$). Then, in terms of the matter variable $\tau_{a}=\tilde{\rho}_{a}-r^{2}\mathcal{T}v_{a}$ (which we notice it vanishes in vacuum), the equation $\varepsilon^{ab}\nabla_{a}O_{b}=\varepsilon^{ab}\nabla_{a}\Upsilon_{b}$ gives
\begin{equation}\label{EinsteinVec}
r^{2}\nabla_{a}\left(r^{-2}\nabla^{a}\Omega\right)-\frac{\lambda^{2}-2}{r^{2}}\Omega=r^{2}\epsilon^{ab}\left(\nabla_{a}\tau_{b}+\nabla_{a}\Upsilon_{b}\right)\, ,
\end{equation}
which reduces to a master equation in vacuum. Before fixing the matter content to be that of Maxwell's theory, we consider the dipole mode $l=1$. For this mode only two of the equations above exist, 
\begin{align}
&\nabla^{b}\left(r^{4}\nabla_{[a}v_{b]}\right)=r^{2}\left(\tau_{a}+\Upsilon_{a}\right)\, ,\\ \notag \\
&\nabla_{a}\left(r^{2}(\tau^{a}+\Upsilon^{a})\right)=0\, ,
\end{align}
and we recall that $\tau_{a}$ and $\nabla_{[a}v_{b]}$ are gauge invariant (even though $\tilde{\rho}_{a}$ and $\tilde{j}_{a}$ are not gauge invariant for this mode). The second equation implies that there must be a function $\tau$ (defined up to the addition of a constant) satisfying
\begin{equation}\label{tau}
    \nabla_{a}\tau=r^{2}\varepsilon_{ab}\left(\tau^{b}+\Upsilon^{b}\right)
\end{equation}
and the first equation then gives 
\begin{equation}
    \Omega=\tau\, ,
\end{equation}
where $\Omega$ is defined as above. That is, the gravitational degree of freedom becomes non-dynamical for this mode, an is fully given by the matter mode $\tau$, which will be dynamical in general (an example is Maxwell's theory, as we show next). In vacuum, the most general solution is simply $\Omega=\text{constant}$, which corresponds to inducing a small rotation into the hole.

Let us now specify the equations above for Maxwell's theory. The energy-momentum fluctuation $\tau_{a}$ associated to the Maxwell field follows from the last expression in \eqref{emfluct},
\begin{equation}
\tau_{a}=-\frac{i}{r^{2}\lambda^{2}}\left(\varepsilon_{ab}\nabla^{b}\Phi^{-}+4\pi r^{2} \mathcal{J}_{a}^{-}\right)\, .
\end{equation}
Now, in terms of the gravitational variable $\mathfrak{g}^{-}\equiv r^{-1}\left(\Omega+(i/\lambda^{2})\Phi^{-}\right)$ and renaming $\mathfrak{a}^{-}\equiv \Phi^{-}$, the Einstein and Maxwell equations  \eqref{EinsteinVec} and \eqref{MaxM} can be cast in the form
\begin{equation}
\square\left( \begin{array}{@{}c@{}} \mathfrak{g}^{-} \\ \mathfrak{a}^{-}\end{array} \right)+\left(\begin{array}{@{}cc@{}} f^{-}_{\mathfrak{g}\mathfrak{g}} & f^{-}_{\mathfrak{g}\mathfrak{a}} \\  f^{-}_{\mathfrak{a}\mathfrak{g}} &  f^{-}_{\mathfrak{a}\mathfrak{a}}  \\ \end{array}\right)\left( \begin{array}{@{}c@{}} \mathfrak{g}^{-} \\ \mathfrak{a}^{-} \end{array} \right)=\left( \begin{array}{@{}c@{}} S^{-}_{\mathfrak{g}} \\ S^{-}_{\mathfrak{a}} \end{array} \right)
\end{equation}
where $f^{-}_{\mathfrak{g}\mathfrak{g}},f^{-}_{\mathfrak{a}\mathfrak{g}},f^{-}_{\mathfrak{g}\mathfrak{a}}$ and $f^{-}_{\mathfrak{a}\mathfrak{a}}$ are functions of $r$ only (whose particular form is unimportant), and the source terms are 
\begin{align}
S^{-}_{ \mathfrak{g}}&=r\varepsilon^{ab}\left(\nabla_{a}\Upsilon_{b}-\frac{4\pi i}{\lambda^{2}}\nabla_{a}\mathcal{J}_{b}^{-}\right)\, ,\\ 
S^{-}_{\mathfrak{a}}&=-4\pi\left(\lambda^{2}\mathcal{V}^{-}+\varepsilon^{ab}\nabla_{a}\left(r^{2}\mathcal{J}^{-}_{b}\right)\right)\, .
\end{align}
Finally, introducing the parameters 
\begin{equation}
 q_{1}=3M+\Delta\, , \ \ q_{2}=3M-\Delta\, , \ \ \Delta=\sqrt{9 M^{2}+4\bar{\bold{C}}\bold{C}(\lambda^{2}-2)}\, ,
\end{equation}
the equations can be decoupled trading $(\mathfrak{g}^{-},\mathfrak{a}^{-})$ in favour of two variables $(\Psi^{-}_{1},\Psi^{-}_{2})$ defined by 
\begin{equation}
\left( \begin{array}{@{}c@{}} \mathfrak{g}^{-} \\  \mathfrak{a}^{-} \end{array} \right)=\mathcal{A}^{-}\left(\begin{array}{@{}cc@{}} 1 & \alpha^{-}  \\  \frac{i\lambda^{2}}{\lambda^{2}-2}q_{1} &  \alpha^{-} \frac{i\lambda^{2}}{\lambda^{2}-2} q_{2} \\ \end{array}\right)\left( \begin{array}{@{}c@{}} \Psi^{-}_{1} \\ \Psi^{-}_{2} \end{array} \right)
\end{equation}
where $\mathcal{A}^{-},\alpha^{-}$ are arbitrary non-zero constants. The final decoupled master equations are
\begin{equation}
\left(\square -V_{1,2}^{-} \right)\Psi_{1,2}^{-}=\bold{S}^{-}_{1,2}
\end{equation}
where
\begin{align}
V_{1,2}^{-}&=r^{-4}\left[\lambda^{2} r^{2}-q_{2,1}r+4\bar{\bold{C}}\bold{C}\right]\, ,\\
\bold{S}_{1}^{-}&=\frac{-1}{2\mathcal{A}^{-}\Delta}\left(q_{2} S_{\mathfrak{g}}^{-}+i\frac{\lambda^{2}-2}{\lambda^{2}}S_{\mathfrak{a}}^{-}\right)\, , \\
\bold{S}_{2}^{-}&=\frac{1}{2\mathcal{A}^{-}\Delta\alpha^{-}}\left(q_{1}S_{\mathfrak{g}}^{-}+i\frac{\lambda^{2}-2}{\lambda^{2}}S_{\mathfrak{a}}^{-}\right)\, .
\end{align}
Alternatively, introducing the function
\begin{equation}\label{SUpot}
W_{1,2}(r)=\frac{f(r)}{r\left((\lambda^{2}-2)r+q_{2,1}\right)}
\end{equation}
the potentials can be written as
\begin{equation}\label{potmin}
V^{-}_{1,2}=-q_{2,1}\frac{\text{d}}{\text{d}r}W_{1,2}+q_{2,1}^{2}f^{-1}W^{2}_{1,2}+\lambda^{2}\left(\lambda^{2}-2\right)f^{-1}W_{1,2}\, ,
\end{equation}
as reported in the main text. Regarding the special mode $l=1$, we need to find a $\tau$ satisfying \eqref{tau}. From the last expression in \eqref{emfluct} we have
\begin{equation}\label{eqtau}
    r^{2}\varepsilon_{a}^{\ b}\left(\tau_{b}+2\pi i\mathcal{J}_{b}^{-}\right)=-\frac{i}{2}\nabla_{a}\Phi^{-}\, .
\end{equation}
Now, a relation between $\Upsilon_{a}$ and $\mathcal{J}^{-}_{a}$ follows from the conservation laws \eqref{consC1}-\eqref{consC2} that the external current and the energy-momentum tensor need to satisfy on-shell for consistency. In particular, we get
\begin{align}
    \star d\star \bold{J}=0&\longrightarrow 2\mathcal{J}=\nabla^{a}\left(r^{2}\mathcal{J}_{a}\right)\, , \\ \notag \\
  \hat{\nabla}^{\mu}S_{\mu\nu}=-4\pi i\left(\bar{\bold{J}}^{\alpha}\bold{F}_{\nu\alpha}-\bold{J}^{\alpha}\bar{\bold{F}}_{\nu\alpha}\right)&\longrightarrow \nabla_{a}\left(r^{2}\Upsilon^{a}\right)=4\pi i \mathcal{J}^{-}\, .
\end{align}
Combining these two equations, it follows that there must exist a potential $\omega$ associated to the sources (and defined up to the addition of a constant) that satisfies
\begin{equation}
\nabla_{a}\omega=r^{2}\varepsilon_{a}^{\ b}\left(\Upsilon_{b}-2\pi i\mathcal{J}^{-}_{b}\right) \, .
\end{equation}
By combining this with \eqref{eqtau} we get that taking
\begin{equation}
\tau=\omega-(i/2)\Phi^{-}\, 
\end{equation}
the desired equation \eqref{tau} is satisfied. Finally, recalling that $\Omega=\tau$ (a general fact for the $l=1$ odd mode, as we derived above), the Maxwell equation \eqref{MaxM} becomes
\begin{equation}\label{l1}
    \left[\square-\left(\frac{2}{r^{2}}+4\frac{\bar{\bold{C}}\bold{C}}{r^{4}}\right)\right]\Phi^{-}=S^{-}_{\Phi}
\end{equation}
where
\begin{equation}
    S^{-}_{\Phi}=8 i\frac{\bar{\bold{C}}\bold{C}}{r^{4}}\omega-4\pi \left(2\mathcal{V}^{-}+\varepsilon^{ab}\nabla_{a}\left(r^{2}\mathcal{J}_{b}^{-}\right)\right)\, .
\end{equation}
The dynamics of the odd dipole fluctuations $l=1$ is governed entirely by \eqref{l1}, which describes an electromagnetic mode, while the gravitational mode is fixed as $\Omega=\tau$ . We notice that the potential is precisely $V_{1}^{-}$ with $\lambda^{2}=2$ (see \eqref{potmin}), as one would expect. Finally, we point out that $\Phi^{-}$ is gauge-invariant, since it transforms as $\Phi^{-}\to\Phi^{-}+2(\bar{\bold{C}}\bold{C}/r^{2})\xi^{-}$, but $\xi$ is the harmonic component of a \textit{real} vector field, so $\xi^{-}=0$.

\subsection{Generalised Even Sector}

This sector is governed by equations \eqref{even}, with energy-momentum fluctuations given by \eqref{emfluct}, and Maxwell's equation \eqref{MaxP}. Here we shall also make use of the fact that the background has a timelike Killing vector $t^{a}\equiv(\partial_{t})^{a}$, which in the covariant language of Section \ref{S1} can be written as $t_{a}=-\varepsilon_{ab}r^{b}$. We choose to work with the equations
\begin{align}\label{eei}
E_{0}&\equiv -\frac{1}{2}\tilde{h}^{a}_{\ a}-\mathcal{S}=0\, , \\
E_{1}&\equiv t^{a}\left(E_{a}-\tilde{\theta}_{a}-\Sigma_{a}\right) =0\, ,\\
E_{2}&\equiv t^{b}\nabla_{b}\left[r^{a}\left(E_{a}-\tilde{\theta}_{a}-\Sigma_{a}\right)\right] =0\, ,\\
E_{3}&\equiv t^{a}r^{b}\left(E_{ab}-\tilde{\theta}_{ab}-\Sigma_{ab}\right)=0\, ,\\ \label{eef}
E_{4}&\equiv t^{c}\nabla_{c}\left[r^{a}r^{b}\left(E_{ab}-\tilde{\theta}_{ab}-\Sigma_{ab}\right)\right]=0\, ,
\end{align}
and the Maxwell equation \eqref{MaxP} (indeed, if \eqref{eei}-\eqref{eef} hold, then by the Bianchi identity the rest of Einstein's equations hold, too). $E_{0}$ will be used just to replace the trace of $\tilde{h}_{ab}$ in favour of the source term $\mathcal{S}$. Using $E_{1},E_{2},E_{3}$, and in terms of the variable
\begin{equation}
\mathfrak{g}^{+}=\left[t^{a}r^{b}p_{ab}-t^{a}\nabla_{a}\left(\tilde{k}/r\right)\right]/U(r)\, ,
\end{equation}
where $U(r)=\frac{6M+r(\lambda^{2}-2)}{r}-4\frac{\bar{\bold{C}}\bold{C}}{r^{2}}$, equation $E_{4}$ gives a wave equation of the form
\begin{equation}\label{cang1}
\left(\square+f_{\mathfrak{g}\mathfrak{g}}(r)\right)\mathfrak{g}^{+}+f_{\mathfrak{g}\Phi}(r)t^{a}\nabla_{a}\Phi^{+}=S^{+}_{\mathfrak{g}}
\end{equation}
where $f_{\mathfrak{g}\mathfrak{g}}$ and $f_{\mathfrak{g}\Phi}$ are functions of $r$ whose form is unimportant, and the source term is given by
\begin{align}\notag
S^{+}_{\mathfrak{g}}=&S^{(1)}_{\mathfrak{g}}\ t^{c}\nabla_{c}\left(r^{a}r^{b}\Sigma_{ab}\right)+S^{(2)}_{\mathfrak{g}}\ r^{c}\nabla_{c}\left(t^{a}r^{b}\Sigma_{ab}\right)+S^{(3)}_{\mathfrak{g}}\ t^{a}r^{b}\Sigma_{ab}\\ \notag
&+S^{(4)}_{\mathfrak{g}}\ t^{c}\nabla_{c}\left(r^{a}\Sigma_{a}\right)+S^{(5)}_{\mathfrak{g}}\ r^{c}\nabla_{c}\left(t^{a}\Sigma_{a}\right)+S^{(6)}_{\mathfrak{g}}\ t^{a}\Sigma_{a} \\ \notag
&+S^{(7)}_{\mathfrak{g}}\ t^{c}\nabla_{c}\left(t^{a}\mathcal{J}^{+}_{a}\right)+S^{(8)}_{\mathfrak{g}}\ r^{c}\nabla_{c}\left(r^{a}\mathcal{J}^{+}_{a}\right)+S^{(9)}_{\mathfrak{g}}\ r^{a}\mathcal{J}^{+}_{a}\\ \label{SSg}
&+S^{(10)}_{\mathfrak{g}}\ t^{a}\nabla_{a}\mathcal{S}
\end{align}
where the functions $S^{(i)}_{\mathfrak{g}}$ are given below, in \eqref{Sg}. Now, in terms of the variable 
\begin{equation}
\mathfrak{a}^{+}=t^{a}\nabla_{a}\Phi^{+}+2\lambda^{2}\frac{\bar{\bold{C}}\bold{C}}{r}\mathfrak{g}^{+}
\end{equation}
and with the help of $E_{1}$ and $E_{3}$, all derivatives in the Maxwell equation \eqref{MaxP} are collected in the differential operators $\square \mathfrak{a}^{+}$, and $\square \mathfrak{g}^{+}$. Then, using \eqref{cang1} to eliminate $\square \mathfrak{g}^{+}$ one finds
\begin{equation}\label{canem1}
\left(\square+f_{\mathfrak{a}\mathfrak{a}}(r)\right)\mathfrak{a}^{+}+f_{\mathfrak{a}\mathfrak{g}}(r)\mathfrak{g}^{+}=S^{+}_{\mathfrak{a}}
\end{equation}
where the source term is
\begin{align}\notag
S^{+}_{\mathfrak{a}}=&2\lambda^{2}\frac{\bar{\bold{C}}\bold{C}}{r}S_{\mathfrak{g}}+S^{(1)}_{\mathfrak{a}}t^{a}r^{b}\Sigma_{ab}+S^{(2)}_{\mathfrak{a}}t^{a}\Sigma_{a}+S^{(3)}_{\mathfrak{a}}t^{a}\nabla_{a}\mathcal{S}\\ \label{SSa}
&+S^{(4)}_{\mathfrak{a}}\ r^{a}\mathcal{J}_{a}+S^{(5)}_{\mathfrak{a}}\ t^{c}\nabla_{c}\left[\varepsilon^{ab}\nabla_{a}\left(r^{2}\mathcal{J}_{b}^{+}\right)\right]+S^{(6)}_{\mathfrak{a}}\ t^{a}\nabla_{a}\mathcal{V}^{+}
\end{align}
and the functions $S_{\mathfrak{a}}^{(i)}$ are given below, in \eqref{Sa}. Finally, trading $t^{a}\nabla_{a}\Phi^{+}$ in favour of $\mathfrak{a}^{+}$ and $\mathfrak{g}^{+}$ in \eqref{cang1}, equations \eqref{canem1} and \eqref{cang1} take the form
\begin{equation}
\square\left( \begin{array}{@{}c@{}} \mathfrak{g}^{+} \\ \mathfrak{a}^{+} \end{array} \right)+\left(\begin{array}{@{}cc@{}} \tilde{f}_{\mathfrak{g}\mathfrak{g}} &  \tilde{f}_{\mathfrak{g}\mathfrak{a}}  \\  f_{\mathfrak{a}\mathfrak{g}}  &   f_{\mathfrak{a}\mathfrak{a}}   \\ \end{array}\right)\left( \begin{array}{@{}c@{}} \mathfrak{g}^{+} \\ \mathfrak{a}^{+} \end{array} \right)=\left( \begin{array}{@{}c@{}} S_{\mathfrak{g}}^{+} \\ S_{\mathfrak{a}}^{+} \end{array} \right)
\end{equation}
where again the form of the functions $\tilde{f}_{\mathfrak{g}\mathfrak{g}},\tilde{f}_{\mathfrak{g}\mathfrak{a}}$ is unimportant. As in the odd case, this system of equations can be simply decoupled with a very similar linear transformation, with constant coefficients, trading $(\mathfrak{g}^{+},\mathfrak{a}^{+})$ in favour of $(\Psi^{+}_{1},\Psi^{+}_{2})$ defined by
\begin{equation}
\left( \begin{array}{@{}c@{}} \mathfrak{g}^{+} \\ \mathfrak{a}^{+} \end{array} \right)=\mathcal{A}^{+}\left(\begin{array}{@{}cc@{}} 1 & \alpha^{+}  \\ \frac{\lambda^{2}}{2}q_{1} &  \alpha^{+} \frac{\lambda^{2}}{2}q_{2}  \\ \end{array}\right)\left( \begin{array}{@{}c@{}} \Psi^{+}_{1} \\ \Psi^{+}_{2} \end{array} \right)
\end{equation}
where again $\mathcal{A}^{+}$ and $\alpha^{+}$ are arbitrary non-zero constants. One finally has 
\begin{equation}
\left(\square-V_{1,2}^{+}\right)\Psi_{1,2}^{+}=\bold{S}_{1,2}^{+}
\end{equation}
where
\begin{align}
\bold{S}_{1}^{+}=&\frac{-1}{2  \mathcal{A}^{+} \Delta}\left(q_{2}S^{+}_{\mathfrak{g}}-\frac{2}{\lambda^{2}}S^{+}_{\mathfrak{a}}\right)\, ,\\
\bold{S}_{2}^{+}=&\frac{1}{2  \mathcal{A}^{+} \Delta \alpha^{+}}\left(q_{1}S_{\mathfrak{g}}^{+}-\frac{2}{\lambda^{2}}S^{+}_{\mathfrak{a}}\right)\, ,\\
V^{+}_{1,2}=&q_{2,1}\frac{\text{d}}{\text{d}r}W_{1,2}+q_{2,1}^{2}f^{-1}W^{2}_{1,2}+\lambda^{2}\left(\lambda^{2}-2\right)f^{-1}W_{1,2}
\end{align}
and $W_{1,2}(r)$ is the same as in \eqref{SUpot}. Finally, let us consider the scalar dipolar mode $l=1$. First of all, we notice that while $\Psi^{+}_{2}$ is gauge-dependent, the electromagnetic mode $\Psi^{+}_{1}$ is not. Indeed, for $l=1$ the latter is given by 
\begin{equation}
    \Psi_{1}^{+}=\frac{1}{2\Delta \mathcal{A}^{+}}\left(t^{a}\nabla_{a}\Phi^{+}+4\frac{\bar{\bold{C}}\bold{C}}{r}\left[\left(t^{a}r^{b}p_{ab}-t^{a}\nabla_{a}(\tilde{k}/r)\right)/U(r)\right]\right)\,
\end{equation}
and a gauge transformation of each of its pieces reads
\begin{align}
t^{a}r^{b}p_{ab} &\mapsto t^{a}r^{b}p_{ab}+4rt^{a}r^{b} r_{(a}\nabla_{b)} \left(\frac{\xi}{r^{2}}\right)+2r^{2}t^{a}r^{b}\nabla_{a}\nabla_{b}\left(\frac{\xi}{r^{2}}\right)\, ,\\   
\tilde{k} &\mapsto \tilde{k} + 2\xi +2 r^{3}r^{a}\nabla_{a}\left(\frac{\xi}{r^{2}}\right)\, , \\
\Phi^{+}&\mapsto \Phi^{+}+2\frac{\bar{\bold{C}}\bold{C}}{r^{2}}\xi^{+}=\Phi^{+}+4\frac{\bar{\bold{C}}\bold{C}}{r^{2}}\xi
\end{align}
where we used that $\xi^{+}=2\xi$ since $\xi$ is the harmonic component of a \textit{real} vector field. It is easy to check that this leaves $\Psi_{1}^{+}$ invariant. Now, since for this mode neither $\mathcal{S}$ nor equation $E_{0}$ exist, it is consistent to move to a gauge in which $\tilde{h}^{a}_{\ a}=0$. Then equations $E_{1},E_{2},E_{3},E_{4}$ are identical to the $l\geq2$ case but setting $\lambda^{2}=2$, $\tilde{h}^{a}_{\ a}=\mathcal{S}=0$, and the very same manipulation applies to get the master equations. In fact, it is the equation for $\Psi_{1}^{+}$ alone that governs the entire dynamics of this mode (what could be guessed from the vacuum case, where $\Psi_{2}^{+}$ is pure gauge).
We finally report here the functions $S^{(i)}_{\mathfrak{a},\mathfrak{g}}$ of the source terms \eqref{SSa} and \eqref{SSg},
\begin{equation}\label{Sa}
\begin{aligned}
S^{(1)}_{\mathfrak{a}}=&-\frac{4\lambda^{2}\bar{\bold{C}}\bold{C}}{rU(r)}\, , \\
S^{(2)}_{\mathfrak{a}}=&-\frac{8\lambda^{2}\bar{\bold{C}}\bold{C}\left(3 \bar{\bold{C}}\bold{C}-r\left(6M-3r+r^{3}\Lambda\right)\right)}{3 r^{4}U(r)}\, , \\
S^{(3)}_{\mathfrak{a}}=&-\frac{2\bar{\bold{C}}\bold{C}\lambda^{2}}{r^{2}}\, , \\
S^{(4)}_{\mathfrak{a}}=&\frac{32\pi \bar{\bold{C}}\bold{C}\left(3 \bar{\bold{C}}\bold{C}-r(6M-3r+r^{3}\Lambda)\right)}{3 r^{4}U(r)}\, , \\
S^{(5)}_{\mathfrak{a}}=&-4\pi\, , \\
S^{(6)}_{\mathfrak{a}}=&-4\pi \lambda^{2}\, .
\end{aligned}
\end{equation}
and 
\begin{equation}\label{Sg}
\begin{aligned}
S^{(1)}_{\mathfrak{g}}=&-\frac{r}{f(r)U(r)}\, , \\
S^{(2)}_{\mathfrak{g}}=&\frac{r}{f(r)U(r)}\, , \\
S^{(3)}_{\mathfrak{g}}=&\frac{2 \left(5  \bar{\bold{C}}\bold{C}+3 r^2 f(r)+\Lambda  r^4-3 r^2\right)}{U(r)\left( \bar{\bold{C}}\bold{C}+3 r^2 f(r)+r^2 \left(-\lambda ^2+\Lambda  r^2-1\right)\right)}\, , \\
S^{(4)}_{\mathfrak{g}}=&-\frac{2}{U(r)}\, , \\
S^{(5)}_{\mathfrak{g}}=&\frac{2}{U(r)}\, , \\
S^{(6)}_{\mathfrak{g}}=&-\frac{r^2 f(r) \left(-12  \bar{\bold{C}}\bold{C}-6 r^2 f(r)+r^2 \left(\lambda ^2+4 \Lambda  r^2+4\right)\right)}{r^3 U(r)\left( \bar{\bold{C}}\bold{C}+3 r^2 f(r)+\Lambda  r^4-\left(\lambda ^2+1\right) r^2\right)}\\
&-\frac{\left( \bar{\bold{C}}\bold{C}+\Lambda  r^4-\left(\lambda ^2+1\right) r^2\right) \left(2  \bar{\bold{C}}\bold{C}+r^2 \left(\lambda ^2+2 \Lambda  r^2-2\right)\right)}{r^3 U(r)\left( \bar{\bold{C}}\bold{C}+3 r^2 f(r)+\Lambda  r^4-\left(\lambda ^2+1\right) r^2\right)}\, , \\
S^{(7)}_{\mathfrak{g}}=&\frac{8 \pi }{U(r)\lambda ^2}\, , \\
S^{(8)}_{\mathfrak{g}}=&-\frac{8 \pi }{U(r)\lambda ^2}\, , \\
S^{(9)}_{\mathfrak{g}}=&4\pi\frac{ 2  \bar{\bold{C}}\bold{C}^2+f(r) \left(r^4 \left(\lambda ^2+4 \Lambda  r^2+4\right)-12  \bar{\bold{C}}\bold{C} r^2\right)}{U(r)\lambda ^2 r^3 \left( \bar{\bold{C}}\bold{C}+3 r^2 f(r)+r^2 \left(-\lambda ^2+\Lambda  r^2-1\right)\right)}\\
&+4\pi \frac{\bar{\bold{C}}\bold{C} r^2 \left(-\lambda ^2+4 \Lambda  r^2-4\right)-6 r^4 f(r)^2+r^4 \left(-\lambda ^4+\lambda ^2 \left(1-\Lambda  r^2\right)+2 \left(\Lambda  r^2-1\right)^2\right)}{U(r)\lambda ^2 r^3 \left( \bar{\bold{C}}\bold{C}+3 r^2 f(r)+r^2 \left(-\lambda ^2+\Lambda  r^2-1\right)\right)}\, , \\
S^{(10)}_{\mathfrak{g}}=&\frac{- \bar{\bold{C}}\bold{C}-3 r^2 f(r)+r^2 \left(\lambda ^2-\Lambda  r^2+1\right)}{r^3 U(r)}\, .
\end{aligned}
\end{equation}

\end{document}